 \documentclass[twocolumn]{aastex6}

\usepackage{color}

\newcommand{\bwt}{\begin{widetext}}
\newcommand{\ewt}{\end{widetext}}
\newcommand{\beq}{\begin{equation}}
\newcommand{\eeq}{\end{equation}}
\newcommand{\bea}{\begin{eqnarray}}
\newcommand{\eea}{\end{eqnarray}}

\begin{document}
                    
\title{Equation of State for Nucleonic and Hyperonic Neutron Stars \\ with Mass and Radius Constraints}
\author{Laura Tolos}
\affil{Institute of Space Sciences (CSIC-IEEC), Campus Universitat Aut\`onoma de Barcelona, \\Carrer de Can Magrans, s/n, 08193 Cerdanyola del Vall\`es,
Spain\\
Frankfurt Institute for Advanced Studies,  Goethe University Frankfurt, \\ Ruth-Moufang-Str. 1,
60438 Frankfurt am Main, Germany }
\author{Mario Centelles and Angels Ramos}
\affil{Departament de F\'{\i}sica Qu\`antica i Astrof\'{\i}sica and Institut de Ci\`encies del Cosmos (ICCUB),\\ Universitat de Barcelona,
Mart\'{\i} i Franqu\`es 1, 08028 Barcelona, Spain}

\begin{abstract}
We obtain a new equation of state for the nucleonic and hyperonic inner core of 
neutron stars that fulfills the 2$M_{\odot}$ observations as well as the recent 
determinations of stellar radii below 13 km. The nucleonic 
equation of state is obtained from a new parametrization of the FSU2 relativistic 
mean-field functional that satisfies these latest astrophysical constraints and, 
at the same time, reproduces the properties of nuclear matter and finite nuclei 
while fulfilling the restrictions on high-density matter deduced from heavy-ion 
collisions. On the one hand, the equation of state  of neutron star matter is softened 
around saturation density, which increases the compactness of canonical neutron 
stars leading to stellar radii below 13 km. On the other hand, the equation of 
state is stiff enough at higher densities to fulfill the 2$M_{\odot}$ limit. By a 
slight modification of the parametrization, we also find that the constraints of 
2$M_{\odot}$ neutron stars with radii around 13 km are satisfied when hyperons 
are considered. The inclusion of the high magnetic fields present in magnetars 
further stiffens the equation of state. Hyperonic magnetars with magnetic fields 
in the surface of $ \sim 10^{15}$ G and with values of $\sim 10^{18}$ G in the 
interior can reach maximum masses of 2$M_{\odot}$ with radii in the  12-13 km 
range. 
\end{abstract}


\vspace*{5mm}
\section{Introduction}
\label{sec:intro}

Neutron stars are the most compact known objects without event horizons. They are formed in the aftermath of core-collapse supernovae and are usually observed as pulsars.  Their features, such as the mass and radius, strongly depend on the properties of their dense interior. Thus, neutron stars serve as a unique laboratory for dense matter physics.

With more than 2000 pulsars known up to date, one of the best determined pulsar masses is that of the Hulse-Taylor pulsar of 1.4$M_{\odot}$ \citep{Hulse:1974eb}. Until very recently, the most precise measurements of neutron star masses clustered around this canonical value. Higher masses in neutron star binary systems have been measured in recent years with very high precision, using post-Keplerian parameters. This is the case of the binary milisecond pulsar PSR J1614-2230 of $M=1.97 \pm 0.04 M_{\odot}$ \citep{Demorest:2010bx} and the PSR J0348+0432 of $M=2.01\pm 0.04 M_{\odot}$ \citep{Antoniadis:2013pzd}.  

While the measurement of neutron star masses is 
accurate, the observational determination of their radii is more difficult and, 
as a consequence, comparably accurate values of radii do not yet exist.
The radius of a neutron star can be extracted from the analysis of X-ray spectra 
emitted by the neutron star atmosphere. The 
modeling of the X-ray emission strongly depends on the distance to the source, its 
magnetic field and the composition of its atmosphere, thus making the determination 
of the radius a difficult task. 
As a result, different values for the stellar radii 
have been derived 
\citep{Verbiest:2008gy,Ozel:2010fw,Suleimanov:2010th,Lattimer:2012xj,Steiner:2012xt, 
Bogdanov:2012md,Guver:2013xa,Guillot:2013wu,Lattimer:2013hma,Poutanen:2014xqa, 
Heinke:2014xaa,Guillot:2014lla,Ozel:2015fia,Ozel:2015gia,Ozel:2016oaf,
Lattimer:2015nhk}. In general, the extractions based on the spectral analysis of X-ray emission from quiescent 
X-ray transients in low-mass binaries (QLMXBs) favor small stellar radii in the 9-12 km range, whereas the determinations from neutron stars with recurring powerful bursts may lead to larger radii, of up to 16 km, although they are subject to larger uncertainties and controversy (see the discussion in the analysis of Ref.~\citep{Fortin:2014mya}).
The very recent work of 
Ref.~\citep{Lattimer:2015nhk} indicates that the realistic range for radii of 
canonical neutron stars should be 10.7 km to 13.1 km. This analysis is based on 
observations of  pulsar masses and estimates of symmetry properties derived from neutron matter studies 
and nuclear experiments.
It is expected that robust observational upper 
bounds on stellar radii will be within reach in a near future. With space missions such as 
NICER (Neutron star Interior Composition ExploreR) \citep{2014SPIE.9144E..20A}, 
high-precision X-ray astronomy will be able to offer precise measurements of 
masses and radii \citep{Watts:2016uzu}, while gravitational-wave signals from 
neutron-star mergers hold promise to determine neutron-star radii with a precision 
of 1 km \citep{Bauswein:2011tp,Lackey:2014fwa}.

In anticipation that these upcoming astrophysical determinations could confirm small neutron star sizes, it is important and timely to explore the 
smallest radii that can be delivered by the theoretical models of compressed matter that are able to fulfill the $2 M_\odot$ maximum mass constraint, while reproducing at the same time the phenomenology of atomic nuclei.
The masses and radii of neutron stars are linked to the physics of their interior, 
that is, the equation of state (EoS) of dense matter 
\citep{Lattimer:2004pg,Lattimer:2006xb,Oertel:2016bki}. Many of the current nuclear models for the 
EoS are able to satisfy the 2$M_{\odot}$ 
constraint required by the discovery of massive neutron stars 
\citep{Demorest:2010bx,Antoniadis:2013pzd}. However, the possible existence of 
neutron stars with small radii suggested by recent astrophysical analyses 
\citep{Guillot:2013wu,Guver:2013xa,Lattimer:2013hma,Heinke:2014xaa,Guillot:2014lla,
Ozel:2015fia,Ozel:2016oaf,Lattimer:2015nhk} poses a difficult challenge 
to most of the nuclear models \citep{Dexheimer:2015qha,Jiang:2015bea,Chen:2015zpa,Ozel:2016oaf}. 

 A small 
neutron star radius for a canonical neutron star requires a certain softening of 
the pressure of neutron matter, and hence of the nuclear symmetry energy, around 
1-2 times saturation density $n_0$ ($n_0 \approx 0.16$~${\rm fm}^{-3}$) 
\citep{Lattimer:2006xb,Tsang:2012se,Ozel:2016oaf}.
The star radius could also be reduced by decreasing the pressure of 
the isospin-symmetric part of the EoS in the intermediate density region, but this 
is only possible with severe limitations due to the 
saturation properties of nuclear matter and the constraints on the EoS of dense 
nuclear matter extracted from nuclear collective flow  \citep{Danielewicz:2002pu} 
and kaon production \citep{Fuchs:2000kp,Lynch:2009vc} in high-energy heavy-ion 
collisions (HICs). Moreover, the requirement of maximum masses of 2$M_{\odot}$ does 
not allow a significant reduction of the total pressure.
Indeed, very few models seem to exist that can meet both constraints (small 
radius and large mass) simultaneously, and fewer such models can in addition 
render accurate descriptions of the finite nuclei properties 
\citep{Jiang:2015bea,Horowitz:2000xj,Horowitz:2001ya,Chen:2015zpa,Sharma:2015bna}.

It has also been long known that the transition from nuclear matter to hyperonic 
matter is energetically favored as the density increases inside neutron stars 
\citep{1960AZh....37..193A}. The opening of hyperon degrees of freedom leads to a 
considerable softening of the EoS \citep{Glendenning:1982nc}. As a consequence, the 
maximum neutron star masses obtained are usually smaller than the 2$M_{\odot}$ 
observations \citep{Demorest:2010bx,Antoniadis:2013pzd}. The solution of this 
so-called ``hyperon puzzle'' is not easy, and requires a mechanism that could provide 
additional repulsion to make the EoS stiffer. Possible mechanisms could be: 
1) stiffer hyperon-nucleon and/or hyperon-hyperon interactions, see  the recent works 
\citep{Bednarek:2011gd,Weissenborn:2011ut,Oertel:2014qza,Maslov:2015msa}; 2) 
inclusion of three-body forces with one or more hyperons, see 
\citep{Vidana:2010ip,Yamamoto:2014jga,Lonardoni:2014bwa} for recent studies; 3) 
appearance of other hadronic degrees of freedom such as the $\Delta$ isobar 
\citep{Drago:2014oja} or meson condensates that push the onset of hyperons to higher 
densities; and 4) appearance of a phase transition to deconfined quark matter at 
densities below the hyperon threshold 
\citep{Alford:2006vz,Zdunik:2012dj,Klahn:2013kga}.
For a detailed review on the ``hyperon puzzle'', we refer the reader to 
Ref.~\citep{Chatterjee:2015pua} and references therein.

The presence of strong magnetic fields inside neutron stars is another 
possible source for a stiffer EoS that could sustain masses of 2$M_{\odot}$.
Anomalous X-ray pulsars and soft $\gamma$-ray repeaters are  
identified with highly magnetized neutron stars with a surface magnetic field of $ 
\sim 10^{14}-10^{15}$ G \citep{Vasisht:1997je, Kouveliotou:1998ze,Woods:1999wa}. 
This class of compact objects has been named ``magnetars", i.e. neutron stars with 
magnetic fields several orders of magnitude larger than the canonical surface dipole 
magnetic fields B $ \sim 10^{12}-10^{13}$ G of the bulk of the pulsar population 
\citep{Mereghetti:2008je,Rea:2011fa,Turolla:2015mwa}. It has been shown that the 
magnetic fields larger than $B/B_c^e=10^5$, with $B_c^e=4.414 \times 10^{13}$ G  
being the critical magnetic field at which the electron cyclotron energy is equal to 
the electron mass, will affect the EoS of dense nuclear matter 
\citep{Chakrabarty:1997ef,Bandyopadhyay:1998aq,Broderick:2000pe,Suh:2000ni,
Harding:2006qn,Chen:2005dx,Rabhi:2008je,Dexheimer:2011pz,Strickland:2012vu}.
The study of the effects upon the EoS of hyperonic matter of very strong 
magnetic fields ($B \sim 10^{18}$--$10^{19}$ G in the star center) was initiated 
in Ref.~\citep{Broderick:2001qw} and has been recently addressed in 
\citep{Rabhi:2009ii,Sinha:2010fm,Lopes:2012nf,Gomes:2014dka}. 


In the present paper we reconcile the $2 M_{\odot}$ mass observations with the 
recent analyses of radii below 13 km for neutron stars, while fulfilling the 
constraints from the properties of nuclear matter, nuclei and HICs at high energy. 
This is accomplished for neutron stars with nucleonic and hyperonic cores. 
The formalism is based on the covariant field-theoretical approach to hadronic 
matter (see for example \citep{Serot:1984ey,Serot:1997xg}, chapter~4 of \citep{Glendenning:2000},
and references therein).
The nucleonic EoS is obtained as a new parameterization of the 
nonlinear realization of the relativistic mean-field (RMF) model 
\citep{Serot:1984ey,Serot:1997xg,Glendenning:2000,Chen:2014sca}.
Starting from the recent RMF parameter set FSU2 \citep{Chen:2014sca}, we find that
by softening the pressure of neutron star matter in the neighborhood of saturation one can 
accommodate smaller stellar radii, while the properties of nuclear 
matter and finite nuclei are still fulfilled. Moreover, we are able to keep the 
pressure at high densities in agreement with HIC data and sufficiently stiff such 
that it can sustain neutron stars of $\approx\!2 M_\odot$. We denote the new parametrization 
by FSU2R. Next we introduce hyperons in our calculation and fit the 
hyperon couplings to the value of the hyperon-nucleon and hyperon-hyperon optical 
potentials extracted from the available data on hypernuclei. 
Whereas the radius of the neutron stars is insensitive to the appearance of the 
hyperons, we find a reduction of the maximum mass of the neutron stars due to the 
expected softening of the EoS.
However, we find that the 2$M_{\odot}$ constraint is still fulfilled when 
hyperons are considered by means of a slight modification of the parameters of the 
model, denoted as FSU2H, compatible with the astrophysical observations and 
empirical data. 
We also analyze the effect of strong magnetic fields in the mass and radius of  
neutron stars.  
The origin of the intense magnetic fields in magnetars is still open to debate 
and the strength of the inner values is still unknown \citep{Thompson:1993hn,Ardeljan:2004fq,Vink:2006ss}. 
Nevertheless, it is worth exploring the modification
on the EoS and on the neutron star properties induced by magnetic fields that are as large
as the upper limit imposed by the scalar virial theorem \citep{Chandrasekhar:1953zz,Shapiro:1983du}, which is of the order
of $B\sim 2\times 10^8 (M/M_\odot)(R/R_\odot)^{-2}$. 
For a star of $R\sim 10$ km and $M\sim 2M_\odot$ the magnetic field could then reach around $2\times 10^{18}$ G. 
In our study we have magnetic fields close to this value only at the very center of the star and assume a magnetic field profile toward a value of $10^{15}$ G at the surface, 
hence fulfilling the stability constraint. From the calculations with the proposed EoS we 
conclude that nucleonic and hyperonic magnetars with a surface 
magnetic field of $ \sim 10^{15}$ G and with magnetic fields values of $\sim 
10^{18}$ G in the interior can reach maximum masses of  2$M_{\odot}$ with 
radii in the 12-13 km range. 

The paper is organized as follows. In Sec.~\ref{sec:formalism} we present the
RMF model and the inclusion of magnetic fields for the 
determination of the EoS in beta-equilibrated matter. In Sec.~\ref{sec:calibration} 
we show how 
we calibrate the nucleonic model FSU2R by fulfilling the constraints of  $2 
M_{\odot}$ mass observations and small neutron star radii, as well as the 
properties 
of nuclear matter, nuclei and HICs at high energy. Then, in 
Sec.~\ref{sec:hyperon} we introduce hyperons and magnetic fields and provide a 
slightly changed parametrization, FSU2H, that also fulfills the observational and 
experimental requirements while allowing for maximum masses of $2 M_{\odot}$. We 
finally summarize our results in Sec.~\ref{sec:summary}.

\vspace*{5mm}
\section{Formalism}
\label{sec:formalism}

Our starting point is the  RMF model of matter, where baryons interact through the exchange of 
mesons and which provides a covariant description of the EoS and nuclear systems. 
The Lagrangian density of the theory can be written as 
\citep{Serot:1984ey,Serot:1997xg,Glendenning:2000,Chen:2014sca}
\beq
{\cal L}= \sum_{b}{\cal L}_{b} + {\cal L}_{m}+ \sum_{l=e, \mu}{\cal L}_{l} ,
\label{lan}
\eeq
with the baryon ($b$), lepton ($l$=$e$, $\mu$), and meson ($m$=$\sigma$, $\omega$, 
$\rho$ and $\phi$) Lagrangians given by
\bea
{\cal L}_{b}&=&\bar{\Psi}_{b}(i\gamma_{\mu}\partial^{\mu}-q_{b}\gamma_{\mu}A^{\mu}- 
m_{b}  \nonumber \\
&+&  g_{\sigma b}\sigma
-g_{\omega b}\gamma_{\mu}\omega^{\mu}-g_{\phi b}\gamma_{\mu}\phi^{\mu}-g_{\rho b}\gamma_{\mu}\vec{I}_b \, \vec{\rho \, }^{\mu} 
)\Psi_{b} , \nonumber \\[2mm] 
{\cal L}_{l}&=& \bar{\psi}_{l}\left(i\gamma_{\mu}\partial^{\mu}-q_{l}\gamma_{\mu}A^{\mu}
-m_{l}\right )\psi_{l} ,\nonumber \\[2mm] 
{\cal L}_{m}&=&\frac{1}{2}\partial_{\mu}\sigma \partial^{\mu}\sigma
-\frac{1}{2}m^{2}_{\sigma}\sigma^{2} - \frac{\kappa}{3!} (g_{\sigma N}\sigma)^3 - \frac{\lambda}{4!} (g_{\sigma N}\sigma)^4 \nonumber \\
&-& \frac{1}{4}\Omega^{\mu \nu} \Omega_{\mu \nu} +\frac{1}{2}m^{2}_{\omega}\omega_{\mu}\omega^{\mu}  + \frac{\zeta}{4!}   (g_{\omega N}\omega_{\mu} \omega^{\mu})^4 \nonumber \\
&-&\frac{1}{4}  \vec{R}^{\mu \nu}\vec{R}_{\mu \nu}+\frac{1}{2}m^{2}_{\rho}\vec{\rho}_{\mu}\vec{\rho \, }^{\mu} + \Lambda_{\omega} g_{\rho N}^2 \vec{\rho}_{\mu}\vec{\rho \,}^{\mu} g_{\omega N}^2 \omega_{\mu} \omega^{\mu} \nonumber \\
&-&\frac{1}{4}  P^{\mu \nu}P_{\mu \nu}+\frac{1}{2}m^{2}_{\phi}\phi_{\mu}\phi^{\mu} -\frac{1}{4} F^{\mu \nu}F_{\mu \nu} ,
\label{lagran}
\eea
where $\Psi_{b}$ and $\psi_{l} $ are the baryon and lepton Dirac fields, respectively.  The mesonic and electromagnetic field strength tensors are $\Omega_{\mu \nu}=\partial_{\mu}\omega_{\nu}-\partial_{\nu}\omega_{\mu}$, $\vec{R}_{\mu 
\nu}=\partial_{\mu}\vec{\rho}_{\nu}-\partial_{\nu}\vec{\rho}_{\mu} $, $P_{\mu \nu}=\partial_{\mu}\phi_{\nu}-\partial_{\nu}\phi_{\mu}$ and  $F_{\mu \nu}=\partial_{\mu}A_{\nu}-\partial_{\nu}A_{\mu}$. 
The electromagnetic field is assumed to be externally generated, and, as we will discuss below,
we do not consider the coupling of the particles to the electromagnetic field tensor via the baryon anomalous magnetic moments.
The strong interaction couplings of a meson to a certain baryon are denoted by $g$ (with $N$ indicating nucleon),  the electromagnetic couplings by $q$ and the baryon, meson and lepton masses by $m$.  The vector $\vec{I}_b$ stands for the isospin operator.

The Lagrangian density (\ref{lagran}) incorporates scalar and vector meson 
self-interactions as well as a mixed quartic vector meson interaction. The nonlinear 
meson interactions are important for a quantitative description of nuclear matter 
and finite nuclei, as they lead to additional density dependence that represents in 
an effective way the medium dependence induced by many-body correlations. 
The scalar self-interactions with coupling constants $\kappa$ and $\lambda$,
introduced by \citep{Boguta:1977xi}, are responsible for 
softening the EoS of symmetric nuclear matter around saturation density and allow 
one to obtain a realistic value for the compression modulus of nuclear matter 
\citep{Boguta:1977xi,Boguta:1981px}. The quartic isoscalar-vector self-interaction 
(with coupling $\zeta$) softens the EoS at high densities \citep{Mueller:1996pm}, 
while the mixed quartic isovector-vector interaction (with coupling 
$\Lambda_{\omega}$) is introduced \citep{Horowitz:2000xj,Horowitz:2001ya} to modify 
the density dependence of the  nuclear symmetry energy, which measures the energy 
cost involved in changing the protons into neutrons in nuclear matter.

The Dirac equations for baryons and leptons are given by
\bea
&&(i\gamma_{\mu}\,\partial^{\mu}-q_{b}\,\gamma_{\mu}\,A^{\mu}-m^{*}_{b} \nonumber \\
&&-g_{\omega b}\, \gamma_{0} \, \omega^{0} -g_{\phi b} \,\gamma_{0}\, \phi^{0} 
-g_{\rho b}\, I_{3 b}\, \gamma_{0} \,\rho_3^{0}) \Psi_{b}=0 , \label{MFbary}\\
&&\left(i\gamma_{\mu}\,\partial^{\mu}-q_{l}\,\gamma_{\mu} \,A^{\mu}-m_{l} \right) \psi_{l}=0 , \label{MFlep}
\eea
where the effective baryon masses are defined as
\beq
m^{*}_{b}=m_{b}-g_{\sigma b}\sigma \label{effmass} .
\eeq
The field equations of motion follow from the Euler-Lagrange equations. In the mean-field approximation, the meson fields are replaced by their respective mean-field expectation values, which are given in
uniform matter as  $\bar \sigma= \langle \sigma \rangle$, $\bar\rho=\langle\rho_3^0\rangle$,
$\bar\omega=\langle\omega^0\rangle$ and $\bar \phi=\langle\phi^0\rangle$.  Thus, the equations of motion for the meson fields in the mean-field approximation for the uniform medium are
\begin{eqnarray}
&&m_\sigma^2 \, \bar \sigma + \frac{\kappa}{2} g_{\sigma N}^3 \bar \sigma^2 + \frac{\lambda}{3!}  g_{\sigma N}^4 \bar \sigma^3 = \sum_{b} g_{\sigma b} n_b^s , \label{eqsigma} \\
&& m_\omega^2 \, \bar \omega + \frac{\zeta}{3!}  g_{\omega N}^4 \bar \omega^3 + 2 \Lambda_{\omega} g_{\rho N\,}^2  g_{\omega N}^2  \bar \rho^2 \bar \omega = \sum_{b} g_{\omega b} n_b  , \label{eqomega} \\
&& m_\rho^2 \,  \bar \rho + 2 \Lambda_{\omega} g_{\rho N}^2  g_{\omega N}^2  \bar \omega^2 \bar \rho= \sum_{b} g_{\rho b} I_{3 b} n_b , \label{eqrho}\\
&& m_\phi^2 \bar \phi \, = \sum_{b} g_{\phi b} n_b ~, \label{eqphi}
\end{eqnarray}
where  $I_{3 b}$ represents the third component of isospin of baryon $b$, with the convention $I_{3 p}=1/2$. The quantities $n_b^s= \langle \bar \Psi_b \Psi_b \rangle$ and $n_b=\langle \bar \Psi_b\gamma^0 \Psi_b\rangle$ are the scalar and baryon density for a given baryon, respectively.

In the presence of a magnetic field, the single-particle energy of the charged baryons and leptons is quantized in the perpendicular direction to the magnetic field. 
Taking  the magnetic field in the $z$-direction, $\vec B=B \hat z$, 
the single particle energies of all baryons and leptons are given by \citep{Broderick:2000pe}
\bea
E^{cb}_{\nu}&=& \sqrt{k^{2}_{z}+ m^{* 2}_{cb}+2\nu |q_{cb}| B
} \nonumber\\
&&+g_{\omega \,  cb} \, \bar \omega+g_{\rho \, cb} \, I_{3 b} \, \bar \rho  +g_{\phi \, cb} \, \bar \phi  \label{enspc1} , \\
E^{ub}&=& \sqrt{k^{2}+ m^{* 2}_{ub}} \nonumber \\
&&+g_{\omega \, ub} \, \bar \omega+  g_{\rho  \,ub} \, I_{3 b}  \, \bar \rho +g_{\phi \, ub} \, \bar \phi  \label{enspc2} , \\
E^{l}_{\nu}&=& \sqrt{k^{2}_{z}+m_{l}^{2}+2\nu |q_{l}| B}\label{enspc3} , 
\eea
with ${cb}$ denoting charged baryons and $ub$ uncharged baryons.  The  quantity
$\nu=\left(n+\frac{1}{2}-\frac{1}{2}\frac{q}{|q|}\sigma_z\right)=0, 1, 2, \ldots$, with $n$ being the principal quantum number and $\sigma_z$ the Pauli matrix, indicates the Landau levels of the fermions 
with electric charge $q$. 

%
As mentioned above, we have omitted the coupling of the baryons to the electromagnetic field tensor via their anomalous magnetic moments. The interaction of the baryon anomalous magnetic moments with the field strength has been found to partly compensate for the effects on the EoS associated with Landau quantization \citep{Broderick:2000pe}. However, to see some appreciable changes in the EoS and the neutron star composition, intense fields of the order of  $5\times 10^{18}$ G are needed. Moreover, those effects are mostly concentrated at low densities ($\lesssim 2 n_0$) for such a field strength \citep{Broderick:2000pe,Rabhi:2008je}. Therefore, neglecting the effects associated to the anomalous magnetic moments is a  reasonable approximation in the present work since
we consider neutron stars with magnetic fields at the core of at most $2\times10^{18}$ G and magnetic field profiles that do not reach $5\times10^{17}$ G in the region $\lesssim 2 n_0$.

The Fermi momenta of the charged baryons, $k^{cb}_{F, \nu}$, uncharged baryons, $k^{ub}_{F}$, and leptons, $k^{l}_{F, \nu}$, are related to the Fermi energies $E^{cb}_{F}$, $E^{ub}_{F}$ and  $E^{l}_{F}$ as
\bea
k^{cb}_{F,\nu}&=&\sqrt{(E^{cb}_{F})^{2}- \left( m^{* 2}_{cb}+2\nu |q_{cb}| B \right)} \ , \cr
k^{ub}_{F}&=&\sqrt{(E^{ub}_{F})^{2}-m^{*2}_{ub}} \  , \cr
k^{l}_{F,\nu}&=& \sqrt{(E^{l}_{F})^{2}-\left(m^{2}_{l}+2\nu |q_{l}| B\right) } \ , 
\eea
while the chemical potentials of baryons and leptons are defined as 
\bea
\mu_{b}&=& E^{b}_{F}+g_{\omega  b} \, \bar \omega+g_{\rho b} \,I_{3 b}   \,\bar \rho +g_{\phi b} \,\bar \phi \ ,  \\
\mu_{l} &=& E^{l}_{F}.
\eea
The  largest value of $\nu$ is obtained by imposing that the square of the Fermi momentum of
 the particle is still positive, i.e. by taking the closest integer from below defined by the ratio
$$\nu_{\rm max}=\left[\frac{(E^l_F)^2-m_l^2}{2 |q_l|\, B}\right],\quad \mbox{leptons}$$
$$\nu_{\rm max}=\left[\frac{(E^{cb}_F)^2-{m_{cb}^*}^2}{2 |q_{cb}|\,B}\right] , \quad \mbox{charged baryons}.$$
With all these ingredients, the scalar and vector densities for baryons and leptons 
are given by~\citep{Broderick:2000pe}
\bea
n^{s}_{cb}&=&\frac{|q_{cb}|Bm^{*}_{cb}}{2\pi^{2}}\sum_{\nu=0}^{\nu_{\rm max}} r_{\nu} \ \ln\frac{k^{cb}_{F,\nu}+E^{cb}_{F}}
{\sqrt{m^{* 2}_{cb}+2\nu |q_{cb}|B}}  \ , \cr
n^{s}_{ub}&=&\frac{m^{*}_{ub}}{2\pi^{2}} \left[E^ {ub}_{F}k^{ub}_{F}-{m}^{*2}_{ub}\ln
\frac{k^{ub}_{F}+E^{ub}_{F}}{m^{*}_{ub}} \right] \ ,  \cr
n_{cb}&=&\frac{|q_{cb}|B}{2\pi^{2}}\sum_{\nu=0}^{\nu_{\rm max}}  r_{\nu}  \ k^{cb}_{F,\nu} \ ,  \cr
n_{ub}&=&\frac{\left(k^{ub}_{F}\right) ^{3}}{3\pi^{2}} \ ,  \cr
n_{l}&=&\frac{|q_{l}|B}{2\pi^{2}}\sum_{\nu=0}^{\nu_{\rm max}} r_{\nu} \ k^{l}_{F,\nu} \ ,
\eea
where $r_{\nu}$ is the degeneracy of the $\nu$ Landau level, which is $1$ for $\nu=0$ and $2$ for $\nu \neq 0$.

We can now obtain the energy density $\varepsilon$ and pressure $P$ of the system.  The energy  density of matter, $\varepsilon_{\rm matt}$, is given by 
\bea
\varepsilon_{\rm matt}&=&\sum_{b} \varepsilon_{b}+\sum_{l}\varepsilon_{l} \nonumber \\ &+&\frac{1}
{2}m^{2}_{\sigma} \bar \sigma^{2}+\frac{1}{2}m^{2}_{\omega} \bar \omega^{2}+\frac{1}{2}m^{2}_{\rho} \bar \rho^{2} +\frac{1}{2}m^{2}_{\phi} \bar \phi^{2}\cr
&+& \frac{\kappa}{3 !} (g_{\sigma} \bar \sigma)^ 3+ \frac{\lambda}{4 !} (g_{\sigma} \bar \sigma)^ 4 \cr
&+& \frac{\zeta}{8} (g_{\omega} \bar \omega)^4 + 3 \Lambda_{\omega} (g_{\rho} g_{\omega} \bar \rho \, \bar \omega) ^2 ,
\eea 
where the energy densities of baryons and leptons have the following expressions
\bea
\varepsilon_{cb}&=&\frac{|q_{cb}| B}{4\pi^ {2}}\sum_{\nu=0}^{\nu_{\rm max}} r_{\nu} \times \nonumber \\
&&\left[k^{cb}_{F,\nu}E^{cb}_{F} + ( m^{* 2}_{cb}+2\nu |q_{cb}|B ) 
\ln\frac{k^{cb}_{F,\nu}+E^{cb}_{F}}{\sqrt{m^{* 2}_{cb}+2\nu |q_{cb}| B}} \right] , \cr
\varepsilon_{ub}&=&\frac{1}{8\pi^ {2}}\bigg[k^{ub}_{F}(E_{F}^{ub})^{3}+(k^{ub}_{F})^{3} E_{F}^{ub} - {m}^{*4}_{ub}\ln\frac{k^{ub}_{F}+E^{ub}_{F}}{{m}^{*}_{ub}}  \bigg] , \cr
\varepsilon_{l}&=&\frac{|q_{l}|B}{4\pi^ {2}}\sum_{\nu=0}^{\nu_{\rm max}} r_{\nu} \times \nonumber \\ &&\left[k^{l}_{F,\nu}E^{l}_{F}
+\left(m^{2}_{l}+2\nu |q_{l}|B\right) 
\ln\frac{k^{l}_{F,\nu}+E^{l}_{F}}{\sqrt{m^{2}_{l}+2\nu |q_{l}| B}} \right] .
\eea
The pressure of  matter, $P_{\rm matt}$, is obtained using  the thermodynamic relation
\beq
P_{\rm matt}=\sum_{i}\mu_{i}n_{i}-\varepsilon_{\rm matt} .
\eeq
While the contribution from the electromagnetic field to the energy density is $B^{2}/8 \pi$, we use the so-called ``chaotic field" prescription for the calculation of the pressure of the system \citep{Menezes:2015mqa}, so that we have
\bea
&&\varepsilon=\varepsilon_{\rm matt}+ \frac{B^{2}}{8 \pi} , \\
&&P= P_{\rm matt} + \frac{B^{2}}{24 \pi} .
\label{press}
\eea

\subsection{Neutron star matter in $\beta$-equilibrium}

In order to determine the structure of neutron stars one needs to obtain the EoS over a wide range of densities. For the inner and outer crust of the star we employ the EoS of Ref.~\citep{Sharma:2015bna}, which has been obtained from microscopic calculations. In the core of neutron stars, we find $\beta$-equilibrated globally neutral, charged matter. Consequently, the chemical potentials, $\mu_i$, and particle densities, $n_i$, satisfy the conditions
\bea
&&\mu_i=b_i \mu_n- q_i \mu_e \ ,\nonumber \\
&&0=\sum_{ cb,l} q_i \, n_i \ ,  \nonumber \\
&& n=\sum_{cb,ub} n_i \ ,
\label{beta-eq}
\eea
with $b_i$ the baryon number and $q_i$ the charge of the particle $i$. These 
relations together with Eqs.~(\ref{MFbary},\ref{MFlep}) and the field equations
(\ref{eqsigma}, \ref{eqomega}, \ref{eqrho}, \ref{eqphi}) for $\sigma$, $\omega$, 
$\rho$ and $\phi$ have to be solved self-consistently for total baryon density $n$ 
in the presence of a magnetic field. In this way, we 
obtain the chemical potential and the corresponding density of each species for a 
given $n$, so that we can determine the energy density and pressure of the neutron 
star matter at each density.

Once the EoS is known, the mass $M$ and the corresponding radius $R$ of the neutron star are obtained from solving the Tolman-Oppenheimer-Volkoff (TOV) equations \citep{Oppenheimer:1939ne} 
\bea
&&\frac{dP(r)}{dr}=-\frac{G}{r^2}\left[\varepsilon(r)+P(r)\right] \times \nonumber \\ &&\left[M(r)+4\pi r^3 P(r)\right]\left[1-\frac{2GM(r)}{r}\right]^{-1}  \ , \nonumber \\
&&\frac{dM(r)}{dr}=4 \pi \varepsilon (r) r^2 \ ,
\label{tov}
\eea
where $r$ is the radial coordinate. To solve these equations one needs to specify the initial conditions, namely the enclosed mass and the pressure at the center of the star, $M(r=0)=0$ and $P(r=0)=P_c$, while the energy density is taken from the assumed EoS. The integration of the TOV equations over the radial coordinate ends when $P(r=R)$=0. 



\vspace*{5mm}
\section{Calibration of the nucleonic model}
\label{sec:calibration}

\subsection{Equation of state, stellar masses and stellar radii}

We start our analysis by defining the baseline model for nuclear matter to compute 
masses and radii of neutron stars. Nuclear models that perform similarly well in 
the description of finite nuclei often extrapolate very differently at high 
densities, as usually no information on the high-density sector of the EoS has 
been incorporated in the fitting of the model. In this work we are interested in a 
model that gives neutron star radii as small as possible and massive enough neutron 
stars, in order to reconcile in a unified formalism the new astrophysical 
indications of small stellar radii and the existence of stars of 2$M_{\odot}$ masses, 
while still meeting the constraints from the nuclear data of terrestrial 
laboratories.

\begin{figure}[t]
\begin{center}
\includegraphics[width=0.45\textwidth]{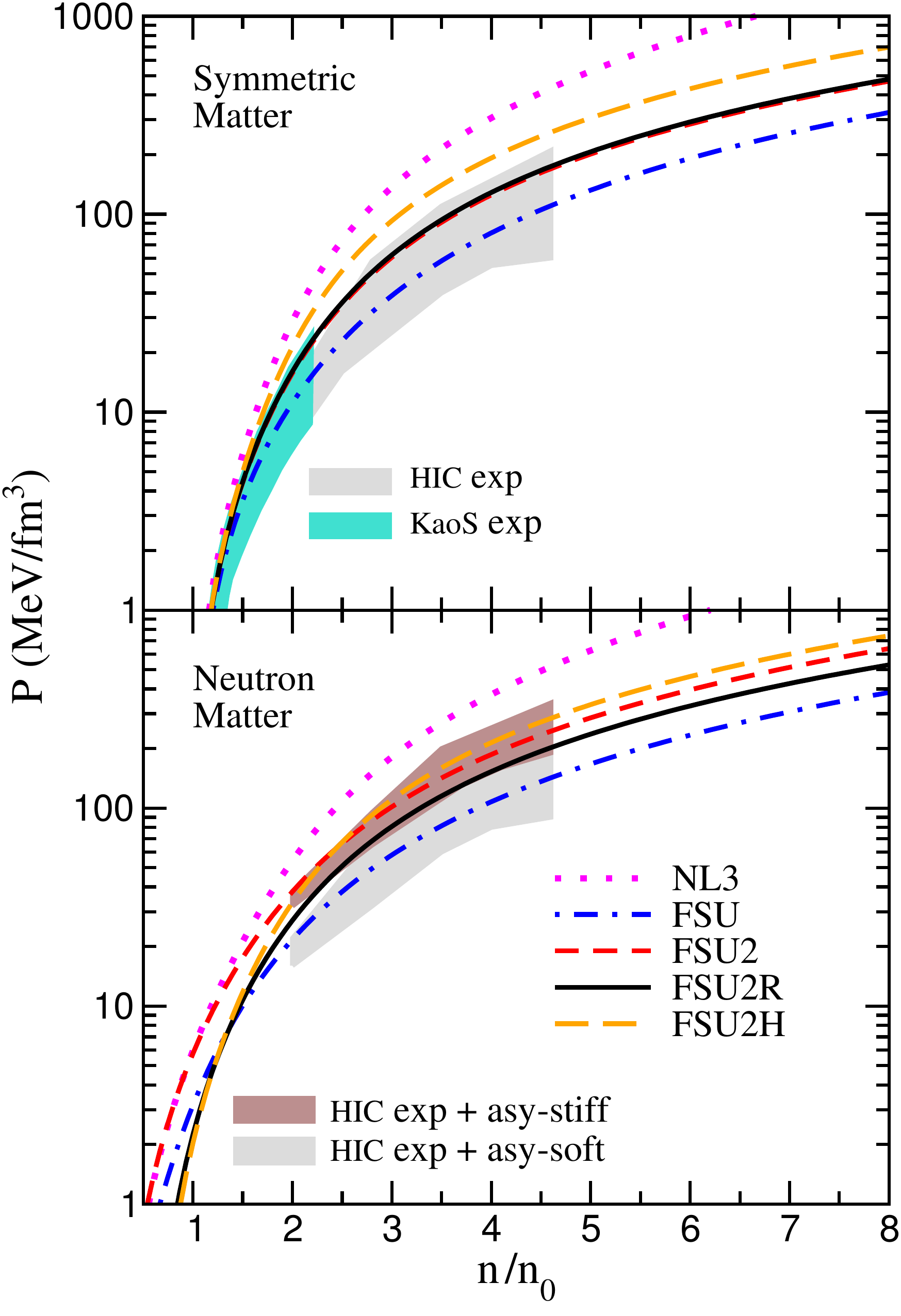}
\caption{Pressure versus baryon density for SNM (upper panel) and PNM (lower panel) 
for the different models presented in the text: NL3 \citep{Lalazissis:1996rd}, FSU 
\citep{ToddRutel:2005zz}, FSU2  \citep{Chen:2014sca}, FSU2R (this work) and FSU2H 
(this work, Sec.~\ref{sec:hyperon}). The regions compatible with the experimental 
data on collective flow \citep{Danielewicz:2002pu} and on kaon production 
\citep{Fuchs:2000kp,Lynch:2009vc} in HICs are depicted in gray and turquoise, 
respectively, in the upper panel. The shaded areas in the panel of PNM correspond to 
the constraints from the flow data supplemented by a soft (gray area) and a stiff 
(brown area) symmetry energy \citep{Danielewicz:2002pu}. }
\label{fig:pressure}
\end{center}
\end{figure}

For the nuclear model we start from the Lagrangian density of 
Eqs.~(\ref{lan},\ref{lagran}) by only considering nucleons and mesons. As mentioned 
in Sec.~\ref{sec:formalism}, the $\zeta$ self-coupling of the $\omega$ meson (cf.\ 
Eq.~(\ref{lagran})) is efficient in softening the EoS at supranormal 
densities while the $\Lambda_{\omega}$ cross coupling of the $\omega$ and $\rho$ 
mesons (Eq.~(\ref{lagran})) regulates the density dependence of the 
symmetry energy. In order to show the effect of these nonlinear contributions to 
the EoS, in Fig.~\ref{fig:pressure} we plot for selected interactions the 
pressures of symmetric nuclear matter (SNM) in the upper panel and of pure neutron 
matter (PNM) in the lower panel. The two shaded 
areas in the SNM panel depict the regions that are compatible with the data on 
collective flow \citep{Danielewicz:2002pu} (gray area) and on kaon production 
\citep{Fuchs:2000kp,Lynch:2009vc} (turquoise region) according to the modeling of 
energetic HICs. The shaded areas in the PNM panel correspond to the constraints 
from the flow data supplemented by a symmetry energy with weak (gray area) or 
strong (brown area) density dependence \citep{Danielewicz:2002pu}.

\begin{table*}[t]
\begin{center}
\begin{tabular}{|c|c|c|c|c|c|}
\hline 
Models & NL3  & FSU & FSU2  & FSU2R  & FSU2H \\ 
\hline
\hline
$m_{\sigma}$ [MeV] & 508.194 & 491.500 &  497.479& 497.479  &  497.479 \\

$m_{\omega}$ [MeV] & 782.501 & 782.500 &  782.500& 782.500  &  782.500 \\

$m_{\rho}$ [MeV] & 763.000& 763.000 &  763.000& 763.000 &  763.000 \\

$g_{\sigma N}^2$ & 104.3871 & 112.1996 &  108.0943& 107.5751 &  102.7200 \\

$g_{\omega N}^2$ & 165.5854 & 204.5469 &  183.7893& 182.3949  &  169.5315\\

$g_{\rho N}^2$ & 79.6000 & 138.4701 & 80.4656& 247.3409  &  247.3409 \\

$\kappa$ & 3.8599 & 1.4203 &  3.0029& 3.0911  &  4.0014 \\

$\lambda$ & $-$0.015905& 0.023762 &  $-$0.000533& $-$0.001680  &  $-$0.013298\\

$\zeta$ & 0.00 & 0.06 &  0.0256& 0.024  &  0.008 \\

$\Lambda_{\omega}$ & 0.00 & 0.03 &  0.000823& 0.05  &  0.05 \\
\hline
$n_0$ $[{\rm fm}^{-3}]$ & 0.1481  & 0.1484  & 0.1505  & 0.1505   & 0.1505 \\

$E/A$ $[{\rm MeV}]$ &  $-$16.24 & $-$16.30 & $-$16.28 & $-$16.28 & $-$16.28 \\

$K$ $[{\rm MeV}]$ & 271.5   & 230.0   &  238.0 &  238.0 &  238.0  \\

$m_N^*/m_N$ & 0.595  & 0.610  & 0.593  & 0.593   & 0.593 \\

$E_{\rm sym}(n_0)$ [MeV] & 37.3& 32.6 &   37.6&  30.2  &  30.2 \\

$L$ [MeV] & 118.2 & 60.5 &  112.8& 44.3  &  41.0 \\

$P (n_0)$ $[{\rm MeV \, fm}^{-3}]$ & 5.99  &  3.18 & 5.81  & 2.27  & 2.06   \\
\hline
 \end{tabular}
\end{center}
\caption{Parameters for the models NL3 \citep{Lalazissis:1996rd}, FSU 
\citep{ToddRutel:2005zz}, FSU2 \citep{Chen:2014sca}, FSU2R (this work) and FSU2H 
(this work, Sec.~\ref{sec:hyperon}). The mass of the nucleon is set to 939 MeV.
Also shown are the corresponding energy per particle ($E/A$), compression modulus 
($K$) and effective nucleon mass $m_N^*/m_N$ at saturation density $n_0$, as well 
as the symmetry energy ($E_{\rm sym}$), slope of the symmetry energy ($L$) and 
 PNM pressure ($P$) at $n_0$.}
\label{t-parameters}
\end{table*}

We first consider the well-known parameter sets NL3 \citep{Lalazissis:1996rd} and 
FSU (also called FSUGold) \citep{ToddRutel:2005zz}. NL3 has $\zeta= 
\Lambda_{\omega}=0$ while FSU has $\zeta=0.06$ and $\Lambda_{\omega}=0.03$ (the 
full set of parameters of the models can be found in Table~\ref{t-parameters}). 
Both NL3 and FSU reproduce quite well a variety of properties of atomic nuclei. 
However, they render two EoSs in SNM with different behavior at supranormal 
densities due to the different $\zeta$ value (we recall that the $\Lambda_{\omega}$ 
coupling does not contribute in SNM). We can see in Fig.~\ref{fig:pressure}(upper 
panel) that above density $n \sim 1.5 \!-\! 2n_0$ the FSU model with $\zeta=0.06$ 
(dot-dashed blue line) yields a much softer SNM pressure than the NL3 model with 
$\zeta=0$ (dotted magenta line). In PNM, the isovector coupling $\Lambda_{\omega}$ 
tunes the change with density of the EoS, as it softens the symmetry energy. 
Indeed, if we compare the same models FSU and NL3 in PNM (see 
Fig.~\ref{fig:pressure}(lower panel)), FSU ($\Lambda_{\omega}=0.03$) has its 
pressure strongly further reduced with respect to NL3 ($\Lambda_{\omega}=0$) in 
the density window from around saturation density $n_0$ up to 
$n \sim 1.5 \!-\! 2n_0$. For densities above $2 n_0$ the softening effect from 
$\Lambda_{\omega}$ is less prominent and the PNM pressures of FSU and NL3 
show comparable differences with the case of SNM. We therefore note, 
consistently with the systematics in earlier works 
\citep{Horowitz:2000xj,Horowitz:2001ya,Carriere:2002bx,Chen:2014sca}, that
the $\Lambda_{\omega}$ and $\zeta$ couplings have a complementary impact on the 
EoS by each one influencing almost separate density sectors.
This will be important for our goals for stellar radii and masses as we shall see 
below.

\begin{figure}[t]
\begin{center}
\includegraphics[width=0.45\textwidth]{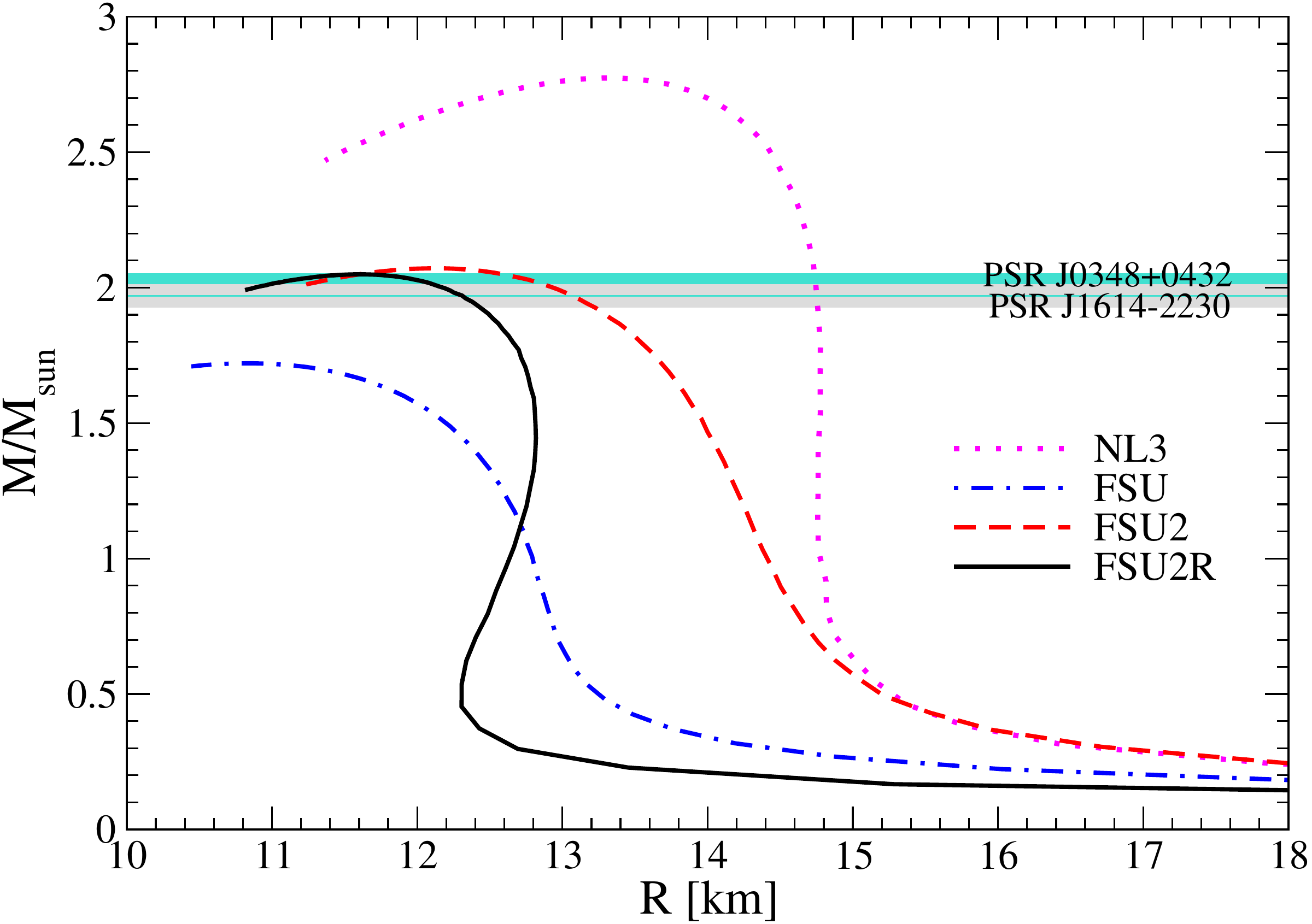}
\caption{Mass versus radius for neutron stars for the models NL3 
\citep{Lalazissis:1996rd}, FSU \citep{ToddRutel:2005zz}, FSU2 \citep{Chen:2014sca} 
and FSU2R (this work). The two shaded bands portray the masses $M=1.97 \pm 
0.04 M_\odot$ in the pulsar PSR J1614--2230  (gray band) \citep{Demorest:2010bx}  
and $M=2.01 \pm 0.04 M_\odot$ in the pulsar PSR J0348+0432 (turquoise band) 
\citep{Antoniadis:2013pzd}. }
\label{fig:mass-radius-nucleons}
\end{center}
\end{figure}

Next, we obtain the mass-radius (M-R) relation of neutron stars for a given EoS by 
solving the TOV equations \citep{Oppenheimer:1939ne}. As mentioned in 
Sec.~\ref{sec:formalism}, for the crust region of the star we have employed the EoS 
recently derived in Ref.~\citep{Sharma:2015bna}.\footnote{We did not find sizable 
changes in our results when we repeated some of the M-R calculations using the 
crustal EoS from the Baym-Pethick-Sutherland model \citep{Baym:1971pw}.}
In this section we focus on neutron stars with cores of purely nucleonic matter;
hence, we compute the EoS of the core assuming a $\beta$-equilibrated and charge 
neutral uniform liquid of neutrons, protons, and leptons (electrons and muons). 
As expected from its stiff EoS, the NL3 set predicts a large maximum 
mass ($M_{\rm max} \approx 2.8 M_\odot$) and large stellar radii ($\approx\,$13~km 
for $M_{\rm max}$ and $\approx\,$15~km for a typical neutron star of $1.5 
M_\odot$), see the M-R relations plotted in Fig.~\ref{fig:mass-radius-nucleons} 
and the values in Table~\ref{tab:starprops}. In comparison, the soft EoS of the
FSU model brings in a dramatic reduction of the stellar masses and radii.
The two shaded bands in Fig.~\ref{fig:pressure} portray the observed masses of the 
heaviest neutron stars known, i.e., $M=1.97 \pm 0.04 M_\odot$ in the 
pulsar PSR J1614--2230 \citep{Demorest:2010bx} and $M=2.01 \pm 0.04 M_\odot$ in the 
pulsar PSR J0348+0432 \citep{Antoniadis:2013pzd}.
These two astrophysical measurements are arguably the most accurate constraints 
available so far to validate or defeat the model predictions for the 
high-density EoS.
\begin{table*}[t]
\begin{center}
\begin{tabular}{|c|c|c|c|c|c|c|c|c|}
\hline
Composition
& Models & $M_{\rm max}/M_\odot$ & $R(M_{\rm max})$& $n_c(M_{\rm max})/n_0$  & 
$R(1.5M_\odot)$  &  $Y$ onset \\
& & & [km] & &  [km] & ($n/n_0$)  \\
\hline
& NL3  & 2.77 & 13.3 &  4.5  & 14.8  & \\
& FSU & 1.72 &10.8 & 7.8  & 12.2  & \\
$pne\mu$& FSU2  &  2.07 & 12.1 & 5.9 & 14.0  & \\
& FSU2R  & 2.05 & 11.6 & 6.3  & 12.8  & \\
& FSU2H  &  2.38  & 12.3 &  5.3  & 13.2  & \\ 
\hline
& NL3  & 2.27 & 12.9 &  5.3  & 14.8  & 1.9 \\
$pnYe\mu$& FSU2  &  1.76 & 12.1 & 6.3 & 13.9 & 2.1\\
& FSU2R  & 1.77 & 11.6 & 6.5  & 12.8 &  2.4\\
& FSU2H  &  2.03  & 12.0 &  5.8 & 13.2 & 2.2\\ 
\hline 
$pne\mu$  & FSU2R & 2.11& 11.6 & 6.1  & 12.8  &  \\
($B_c=2\times 10^{18}$ G)  & FSU2H  &  2.42  & 12.3 &  5.2  & 13.2  & \\ 
\hline
$pnYe\mu$   & FSU2R & 1.88& 11.6 & 6.3 & 12.8 &  2.4\\
($B_c=2\times 10^{18}$ G) & FSU2H  &  2.15  & 12.3 &  5.3  & 13.2& 2.2\\ 
\hline 
 \end{tabular}
\end{center}
\caption{Neutron star properties obtained for the different nuclear models
discussed in this work. Results are shown for nucleonic-only ($pne\mu$) or 
hyperonic ($pnYe\mu$) stars, and including or not a magnetic field having the 
profile of the solid line in Fig.~\ref{fig:B} (Sec.~\ref{sec:hyperon}). The quantity $n_c(M_{\rm max})/n_0$ denotes the central baryonic density at the maximum mass, $M_{\rm max}$, normalized to the corresponding saturation density, $n_0$, whereas  Y onset is the onset of appearance of hyperons normalized to $n_0$. }
\label{tab:starprops}
\end{table*}
The recently formulated relativistic parameter set FSU2 
\citep{Chen:2014sca}---based on the same Lagrangian we are discussing---is one of 
the first best-fit models to take into account the condition of a limiting stellar 
mass of $2 M_\odot$ in the calibration of the parameters (also 
see \citep{Erler:2012qd} and \citep{Chen:2014mza}). The FSU2 
model has been optimized to accurately reproduce the experimental data on a pool 
of properties of finite nuclei with the maximum neutron star mass observable 
included in the fit \citep{Chen:2014sca}. The resulting FSU2 set has $\zeta=0.0256$ 
and $\Lambda_{\omega}=0.0008$, cf.\ Table \ref{t-parameters}. In consonance with 
these values, we can appreciate in Fig.~\ref{fig:pressure} that 
FSU2 predicts an intermediate EoS between the stiff EoS of the NL3 set ($\zeta= 
\Lambda_{\omega}=0$) and the soft EoS of the FSU set ($\zeta=0.06, 
\Lambda_{\omega}=0.03$). Accordingly, FSU2 produces a neutron star mass-radius 
relation located in between the curves of NL3 and FSU in 
Fig.~\ref{fig:mass-radius-nucleons}. FSU2 yields a heaviest stellar mass $M_{\rm 
max}=2.07 M_\odot$ with a radius of 12.1~km, and predicts $1.5 M_\odot$ stars with 
a radius of  14~km, see Table \ref{tab:starprops}.

While the limiting stellar mass is governed by the stiffness of the EoS 
above several times the saturation density $n_0$ (see column $n_c(M_{\rm 
max})/n_0$ in Table~\ref{tab:starprops}), the radius of a canonical neutron star is 
dominated by the density dependence of the EoS of PNM at 1--2 times $n_0$
\citep{Lattimer:2006xb,Ozel:2016oaf}. Thus, 
observational information on masses and radii of neutron stars has 
the potential to uniquely pin down the nuclear EoS in a vast density region. As 
mentioned in the Introduction, several of the recent astrophysical analyses for 
radii \citep{Guillot:2013wu,Guver:2013xa,Lattimer:2013hma,Heinke:2014xaa,
Guillot:2014lla,Ozel:2015fia,Ozel:2016oaf} are converging in the 9--12 km range 
(also see Ref.~\citep{Fortin:2014mya} for a detailed discussion). The review study 
of Ref.~\citep{Lattimer:2015nhk} indicates a similar range around 11--13 km  
for the radii of canonical neutron stars. The possibility that neutron stars have 
these small radii is as exciting as it is deeply challenging for nuclear theory. 
Note that small radii demand a sufficiently soft EoS below twice the saturation 
density, while the observed large masses require that the same EoS must be able to 
evolve into a stiff EoS at high densities. It is therefore timely to explore 
whether such small radii can be obtained by the EoS of the covariant field-theoretical 
Lagrangian (\ref{lan},\ref{lagran}), while fulfilling at the same time the maximum mass constraint of $2 M_\odot$ 
and the phenomenology of the atomic nucleus.

To construct the new EoS we start from the FSU2 model and increase the 
$\Lambda_{\omega}$ coupling. This softens the PNM pressure especially up to 
densities of $1.5 \!-\! 2n_0$. For a given stellar mass there is less 
pressure to balance gravity, thereby leading to a more compact object of smaller 
radius. The increase of $\Lambda_{\omega}$ also produces a certain reduction of the 
PNM pressure in the high-density sector. This may spoil the $2 M_\odot$ maximum 
mass but can be counteracted by a decrease of the strength of the $\zeta$ 
coupling. During the change of the ($\Lambda_{\omega},\zeta$) couplings, we refit
the remaining couplings $g_{\sigma N}, g_{\omega N}, g_{\rho N}, \kappa$, and 
$\lambda$ of the nucleon-meson Lagrangian (\ref{lan},\ref{lagran}) by invoking the 
same saturation properties of FSU2 in SNM
(i.e., same saturation density $n_0$, energy per particle $E/A$, compression modulus 
$K$, and effective nucleon mass $m_N^*$) and a symmetry energy 
$E_{\rm sym}=25.7$~MeV at subsaturation density $n= 0.10$~fm$^{-3}$.
The last condition arises from the fact that the binding energies of atomic nuclei 
constrain the symmetry energy at an average density of nuclei of 
$\sim\!0.10$~fm$^{-3}$ better than the symmetry energy at normal density $n_0$ 
\citep{Horowitz:2000xj,Centelles:2008vu}. We found that under this 
protocol a noteworthy decrease of neutron star radii is achieved with 
$\Lambda_{\omega}=0.05$ and $\zeta=0.024$. We refer to the resulting model as 
FSU2R. The coupling constants and several bulk properties of FSU2R are collected in 
Table~\ref{t-parameters}.

We observe in Fig.~\ref{fig:pressure} that the EoS of the new FSU2R model 
is within the boundaries deduced in the studies of energetic HICs  
\citep{Danielewicz:2002pu,Fuchs:2000kp,Lynch:2009vc}.
It is worth noting that FSU2R features a {\sl soft} PNM EoS at $n \lesssim 1.5 
\!-\! 2 n_0$ and a {\sl stiff} PNM EoS at $n \gtrsim 2 n_0$---apparently a 
necessary condition to satisfy small radii and heavy limiting neutron star 
masses. The reduction of the stellar radii in FSU2R compared with the other 
parametrizations of the Lagrangian (\ref{lan},\ref{lagran}) is very clear 
from Fig.~\ref{fig:mass-radius-nucleons}, also see 
Table~\ref{tab:starprops}. The maximum mass of $2.05 M_\odot$ calculated 
with FSU2R is compatible with the heaviest neutron stars 
\citep{Demorest:2010bx,Antoniadis:2013pzd} and is characterized by a radius of 
 11.6 km. For canonical neutron stars of 1.4--1.5 solar masses, FSU2R predicts radii 
of $\approx\! 12.8$  km, which are more compact than in the other EoSs, cf.\ 
Table~\ref{tab:starprops}. Hence, the smaller radii reproduced by the new model 
point toward the reconciliation between the nuclear EoS, the largest neutron star 
masses \citep{Demorest:2010bx,Antoniadis:2013pzd}, and the recent extractions of 
small neutron star sizes from the astrophysical observations of quiescent low-mass 
X-ray binaries \citep{Guillot:2014lla} and X-ray bursters \citep{Guver:2013xa} (also 
see \citep{Guillot:2013wu,Lattimer:2013hma,Heinke:2014xaa,Ozel:2015fia,
Ozel:2016oaf,Lattimer:2015nhk}). We are only aware of similar models RMF012 and 
RMF016 (also called FSUGarnet) introduced in a recent work \citep{Chen:2014mza}. 
The RMF012 and RMF016 models were fitted with the same procedure of the FSU2 model 
of \citep{Chen:2014sca} but requiring values for the neutron skin thickness of the 
$^{208}$Pb nucleus of, respectively, 0.12 fm and 0.16 fm. As reported in 
\citep{Chen:2015zpa,Chen:2014mza}, the RMF016 model supports $2 M_\odot$ neutron 
stars and leads to a radius of 13 km for a $1.4 M_\odot$ star, similarly to the 
predictions we obtain with our FSU2R model.

\vspace*{5mm}
\subsection{Implications for finite nuclei: \\ symmetry energy, slope of the symmetry energy \\ and neutron skin thickness}
\label{sec:nuclei}

\begin{figure}[t]
\begin{center}
\includegraphics[width=0.45\textwidth]{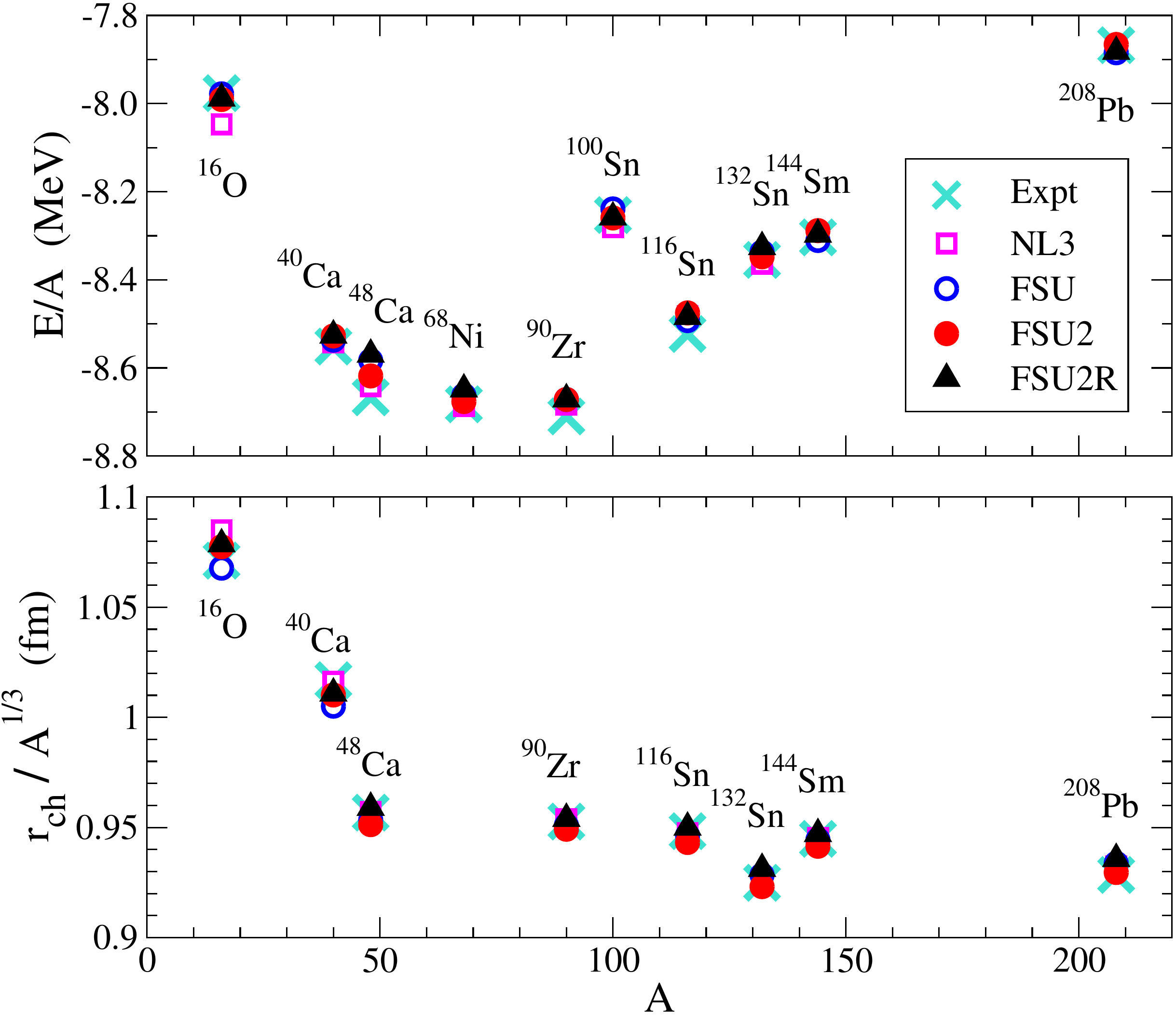}
\caption{Energy per nucleon $E/A$ and charge radius $r_{\rm ch}$ over $A^{1/3}$, 
where $A$ is the mass number, of several nuclei with magic proton and/or neutron 
numbers. The values calculated with the models discussed in the text are compared 
with experiment. The experimental data are from Ref.~\citep{1674-1137-36-12-003} 
for the energies and from Ref.~\citep{Angeli201369} for the charge radii.}
\label{fig:nuclei}
\end{center}
\end{figure}

Once the new EoS has been calibrated for neutron stars, it is important to review 
its implications for the physics regime of atomic nuclei since this regime is 
accessible in laboratory experiments. We first verify that the new model FSU2R 
is able to provide a satisfactory description of the best known properties of 
nuclei, i.e., nuclear ground-state energies and sizes of the nuclear charge 
distributions. We display in Fig.~\ref{fig:nuclei} our results for the energies and 
charge radii of a set of representative nuclei ranging from the light $^{16}$O 
to the heavy $^{208}$Pb. The experimental data of these same nuclei were used 
in the fit of the FSU2 model in Ref.~\citep{Chen:2014sca}. 
In Fig.~\ref{fig:nuclei}, we show the predictions of our FSU2R model
alongside the experimental values and the results from the parameter 
sets NL3, FSU, and FSU2. It can be seen that the four models successfully reproduce 
the energies and charge radii across the mass table. The agreement of FSU2R with 
experiment is overall comparable to the other models. We find that the differences 
between FSU2R and the experimental energies and radii are at the level of 1\% or 
smaller. We mention that we have not drawn error bars of the experimental data in 
Fig.~\ref{fig:nuclei}, because the nuclear masses and charge radii are measured so 
precisely \citep{1674-1137-36-12-003,Angeli201369} that the experimental 
uncertainties cannot be resolved in the plot.

For our purposes, of special relevance is the fact that the neutron density 
distributions and other isospin-sensitive observables of atomic nuclei are 
closely related to the density dependence of the symmetry energy, which in  
FSU2R has been tailored to supply small stellar radii. The 
stiffness of the symmetry energy with density is conventionally characterized 
by its density slope $L$ at the saturation point: $\displaystyle
L = 3 n_0 \Big( \frac{\partial E_{\rm sym}(n)}{\partial n} \Big)_{n_0}$. The $L$ 
parameter and the pressure $P(n_0)$ of  PNM at saturation density 
are related as $P(n_0)= \frac{1}{3} n_0 L$ \citep{Lynch:2009vc,Lattimer:2015nhk}.
The new FSU2R EoS yields $E_{\rm sym}(n_0)= 30.2$~MeV for the symmetry energy 
at saturation and a slope parameter $L= 44.3$~MeV, which corresponds to a 
mildly soft nuclear symmetry energy. The  PNM pressure at saturation is 
$P(n_0)=2.27$~MeV fm$^{-3}$. We have collected these values in 
Table~\ref{t-parameters} along with the results for $E_{\rm sym}(n_0)$, $L$, and 
$P(n_0)$ from the other discussed EoSs---now, large differences can be appreciated 
among the models.

\begin{figure}[t]
\begin{center}
\includegraphics[width=0.45\textwidth]{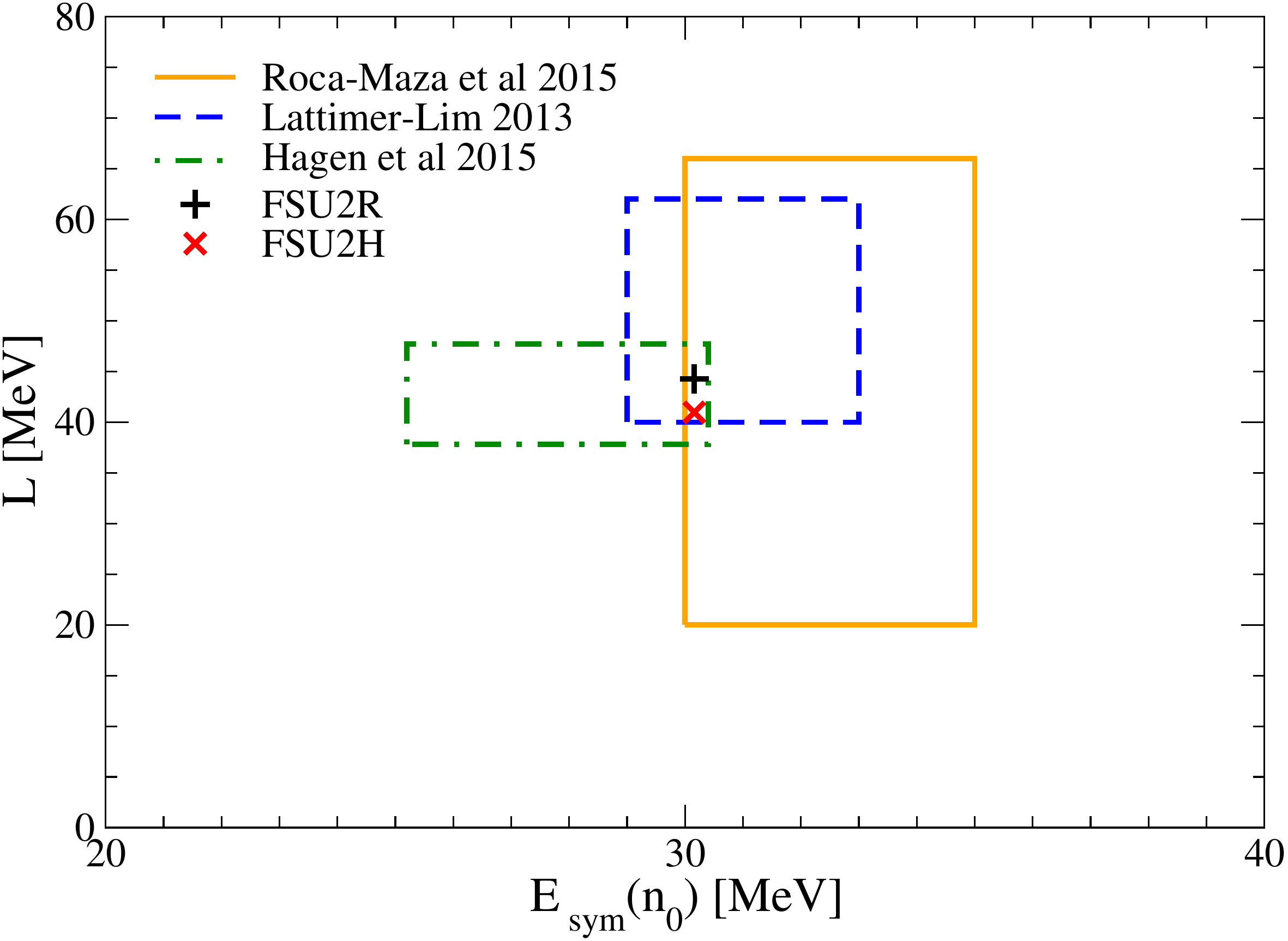}
\caption{Slope of the symmetry energy ($L$) versus symmetry energy ($E_{\rm 
sym}(n_0)$) at saturation for the models FSU2R discussed in this section and FSU2H 
discussed in Sec.~\ref{sec:hyperon}. The rectangular areas are the determinations 
from Refs.~\citep{Lattimer:2012xj,Hagen:2015yea,Roca-Maza:2015eza}.}
\label{fig:esym-l}
\end{center}
\end{figure}

Despite the fact that a precise knowledge of the density dependence of the symmetry energy 
remains elusive, the windows of values for $E_{\rm sym}(n_0)$ and the slope 
parameter $L$ have been continuously narrowed down as the empirical and 
theoretical constraints have improved over recent years (see e.g.\ 
Ref.~\citep{Li:2014oda} for a topical review). Remarkably, the values of 
30.2~MeV for $E_{\rm sym}(n_0)$ and 44.3~MeV for $L$ that we 
find after constraining the EoS to reflect small neutron star radii, turn out to 
be very much consistent with the newest determinations of the symmetry energy and 
its slope at saturation, see Fig.~\ref{fig:esym-l}. Indeed, the quoted FSU2R 
values overlap with the ranges $30 \lesssim E_{\rm sym}(n_0) \lesssim 35$ MeV and 
$20 \lesssim L \lesssim 66$ MeV extracted in Ref.~\citep{Roca-Maza:2015eza} from 
the recent high-resolution measurements at RCNP and GSI of the electric dipole 
polarizability $\alpha_D$ in the nuclei $^{208}$Pb \citep{tamii11}, $^{120}$Sn 
\citep{hashimoto15}, and $^{68}$Ni \citep{rossi13}. We note that the dipole 
polarizability $\alpha_D$, related to the response of nuclei to an external 
electric field, has been identified as one of the strongest isovector indicators 
\citep{reinhard10}. Also note that, compared to hadronic experiments used to probe 
the symmetry energy, the electromagnetic reactions involved in the measurements of 
the $\alpha_D$ observable \citep{tamii11,hashimoto15,rossi13} are particularly 
suited because they are not hindered by large or uncontrolled uncertainties. The 
FSU2R predictions for $E_{\rm sym}(n_0)$ and $L$ also fit within the windows $29 
\lesssim E_{\rm sym}(n_0) \lesssim 33$ MeV and $40 \lesssim L \lesssim 62$ MeV 
obtained in Ref.~\citep{Lattimer:2012xj} from the combined analysis of a variety of 
empirical nuclear constraints and astrophysical informations, which are in line
with similar windows obtained in other recent studies 
\citep{Li:2014oda,Tsang:2012se,Lattimer:2015nhk}.
It also deserves to be mentioned that the $E_{\rm sym}(n_0)$ and $L$ values 
of the FSU2R EoS are quite compatible with the theoretical ranges 
$25.2 \lesssim E_{\rm sym}(n_0) \lesssim 30.4$ MeV and $37.8 \lesssim 
L \lesssim 47.7$ MeV that have been derived from the latest progress in {\sl ab 
initio} calculations of nuclear systems with chiral forces~\citep{Hagen:2015yea}.

The neutron skin thickness $\Delta r_{np}= r_n-r_p$ (difference between the neutron 
and proton matter radii) of a heavy nucleus such as $^{208}$Pb, also provides 
strong sensitivity to the symmetry energy and the pressure of neutron-rich matter 
near saturation \citep{Brown2000,Horowitz:2000xj,Horowitz:2001ya,Centelles:2008vu}. 
Basically the same nuclear pressure that is responsible for determining the radius 
of a canonical neutron star, determines how far neutrons extend out further than 
protons in a nucleus. 
By the same token, models that produce smaller stellar radii are expected to 
predict thinner neutron skins. We find that the FSU2R model, constrained to small 
neutron star radii, predicts $\Delta r_{np}= 0.133$ fm in $^{208}$Pb. 
Unfortunately, neutron skins are difficult to extract from experiments in a 
model-independent fashion. The new experiments to measure neutron 
skins are being designed with electroweak and electromagnetic probes where, 
unlike hadronic experiments, the interactions with the nucleus 
\citep{Abrahamya12}, or at least the initial state 
interactions \citep{Tarbert:2013jze}, are not complicated by the strong force.
The challenging, purely electroweak (nearly model-independent) measurement of the 
neutron skin of $^{208}$Pb by parity violating electron scattering at JLab 
\citep{Abrahamya12,Horowitz:2012tj} has been able to provide $\Delta r_{np} 
= 0.302 \pm 0.177$ fm for this isotope \citep{Horowitz:2012tj}, although the data 
is not conclusive due to the large error bars (a follow-up measurement at JLab with 
better statistics has been proposed). The recent measurement of the neutron 
skin of $^{208}$Pb at the MAMI facility from coherent pion production by photons
\citep{Tarbert:2013jze} has obtained $\Delta r_{np} = 0.15 \pm 0.03$ fm. A similar 
range $0.13 \lesssim \Delta r_{np} \lesssim 0.19$ fm for 
$^{208}$Pb is extracted \citep{Roca-Maza:2015eza} by comparing theory with the 
accurately measured electric dipole polarizability in $^{208}$Pb \citep{tamii11}, 
$^{120}$Sn \citep{hashimoto15}, and $^{68}$Ni \citep{rossi13}. Thus, the FSU2R 
prediction of a neutron skin of 0.133 fm in $^{208}$Pb turns out to be fairly 
compatible within error bars with the recent determinations of this 
isospin-sensitive observable.

In summary, when the nuclear EoS has been constrained to encode the recent 
astrophysical indications of small neutron star radii, yet without compromising 
massive stars, a high degree of consistency has emerged between the predictions 
of the model and the latest terrestrial informations on the symmetry energy, 
its density dependence, and neutron skins, as well as with the constraints 
inferred from state-of-the-art {\sl ab initio} microscopic calculations 
\citep{Hagen:2015yea}. All in all, we believe that the present findings make a 
compelling case in favor of the prospect that neutron stars may have small, or 
moderate-to-small, radii.

\vspace*{5mm}
\section{Hyperons and magnetic field}
\label{sec:hyperon}
Having calibrated the nuclear model to produce small neutron star radii and fulfill maximum masses of ~$2 M_\odot$, while at the same time reproducing the phenomenology of atomic nuclei and the empirical constraints from collective flow and kaon production in HICs, we explore in this section the effect on the EoS and neutron stars of including hyperons and magnetic fields.

We should first determine the value of the hyperon couplings in our  RMF model. Those couplings are calculated by fitting the experimental data available for hypernuclei, in particular, the value of the optical potential of hyperons extracted from these data. In our model,
the contribution to the potential of a hyperon $i$ in $j$-particle matter is given by
\bea
U_i^{(j)}(n_j) &=& - g_{\sigma i} \, \bar \sigma^{(j)} + g_{\omega i} \,  \bar \omega^{(j)} +  g_{\rho i}  \, I_{3 i} \, \bar \rho^{(j)} + g_{\phi i} \, \bar \phi^{(j)},  \  \ \ \ \ \
\label{Ypot}
\eea
where $ \bar \sigma^{(j)}$, $\bar \omega^{(j)}$, $\bar \rho^{(j)}$ and $\bar \phi^{(j)}$ are the values of the meson fields in the $j$-particle matter and $I_{3i}$ stands for the third component of the isospin operator.

The couplings of the hyperons to the vector mesons are related to the nucleon couplings, $g_{\omega N}$ and $g_{\rho N}$, by assuming SU(3)-flavor symmetry, the vector dominance model and ideal mixing for the physical $\omega$ and $\phi$ mesons,  as e.g. employed in many recent works \citep{Schaffner:1995th,Banik:2014qja,Miyatsu:2013hea,Weissenborn:2011ut,Colucci:2013pya}. This amounts to assuming the following relative coupling strengths:
\begin{eqnarray}
g_{\omega \Lambda}:g_{\omega \Sigma}:g_{\omega \Xi}:g_{\omega N}&=&\frac{2}{3}:\frac{2}{3}:\frac{1}{3}:1 \nonumber \\
g_{\rho \Lambda}:g_{\rho \Sigma}:g_{\rho \Xi}:g_{\rho N}&=&0:1:1:1 \nonumber \\
g_{\phi \Lambda}: g_{\phi\Sigma}: g_{\phi\Xi}:g_{\omega N}&=& -\frac{\sqrt{2}}{3}: -\frac{\sqrt{2}}{3}:  -\frac{2\sqrt{2}}{3}: 1 ,  \ \ 
\label{eq:couplings}
\end{eqnarray}
and $g_{\phi N}=0$. Note that the isospin operator $I_{3 i} $ appearing in the definition of the potentials in Eq.~(\ref{Ypot}) implements the relative factor of 2 missing in the 1:1 relation between  $g_{\rho \Sigma}$ and $g_{\rho N}$ displayed in Eq.~(\ref{eq:couplings}), so that the effective coupling of the $\rho$ meson to the $\Sigma$ hyperon ($I_{3}=-1,0,+1$)  is twice that to the nucleon ($I_{3}=-1/2,+1/2$), as required by the symmetries assumed. 

The coupling of each hyperon to the $\sigma$ field is adjusted to reproduce the hyperon potential in SNM derived from hypernuclear observables, see e.g. \citep{Hashimoto:2006aw,Gal:2016boi}. 
The $\Lambda$ binding energy of $\Lambda$-hypernuclei is well reproduced by an attractive Woods-Saxon potential of depth $U_\Lambda^{(N)}(n_0) \sim -28$ MeV \citep{Millener:1988hp}. The analyses of the $(\pi^-, K^+)$ reaction data on medium
to heavy nuclei \citep{Noumi:2001tx} performed in \citep{Harada:2006yj,Kohno:2006iq}
revealed a moderately repulsive $\Sigma$-nuclear potential in the nuclear interior of around $10-40$ MeV, while the fits to $\Sigma^-$ atomic data \citep{Friedman:2007zza} indicate a clear transition from an attractive $\Sigma$ potential in the surface, to a repulsive one in the interior, although the size of the repulsion cannot  be precisely determined. 
As for the strangeness $-2$ systems, the Nagara event \citep{Takahashi:2001nm} and other experiments providing consistency checks established the size of the $\Lambda\Lambda$ interaction to be mildly attractive, $\Delta B_{\Lambda\Lambda} (^6_{\Lambda\Lambda} {\rm He}) = 0.67 ±\pm 0.17$ MeV  \citep{Ahn:2013poa}, while the knowledge obtained for the
$\Xi N$ interaction is more uncertain. Analyses of old emulsion data indicate a sizable attractive $\Xi$-nucleus potential of  $U_\Xi^{(N)}(n_0) = - 24 \pm 4$ MeV \citep{Dover:1982ng}, while the
missing-mass spectra of the $(K ^-, K^+)$ reaction on a $^{12}$C target suggest a milder attraction  of $\gtrsim -20$ MeV \citep{Fukuda:1998bi}  or $\sim -14 \pm 2$ MeV \citep{Khaustov:1999bz}. These values are compatible with the recent analysis of the nuclear emulsion event 
KISO, claiming to have observed a nuclear bound state of the $\Xi^-$-$^{14}$N system with a binding
energy of $4.38\pm 0.25$ MeV \citep{Nakazawa}.
From the above considerations, we fix the hyperon potentials in SNM  to the following values:
\begin{eqnarray}
U_{\Lambda}^{(N)}(n_0)&=&-28~{\rm MeV} \nonumber \\
U_{\Sigma}^{(N)}(n_0)&=& + 30~{\rm MeV} \nonumber \\
U_{\Xi}^{(N)}(n_0)&=&-18~{\rm MeV}  \ ,
\label{eq:pots}
\end{eqnarray}
which allow us to determine the couplings $g_{\sigma \Lambda}$, $g_{\sigma \Sigma}$  and $g_{\sigma \Xi}$, from Eq.~(\ref{Ypot}). 
We finally note
that the coupling of the $\phi$ meson to the $\Lambda$ baryon is reduced by 20\% with respect to its SU(3) value in order to obtain a $\Lambda\Lambda$ bond energy in $\Lambda$ matter at a density $n_\Lambda \simeq n_0/5$ of $\Delta B_{\Lambda\Lambda} (n_0/5) = 0.67$ MeV, thereby reproducing the Nagara event \citep{Ahn:2013poa}. 

Let us comment on the fact that the presence of hyperons in the neutron star interior and their influence on the EoS suffer from uncertainties tied to our lack of knowledge of the hyperon-nucleon and hyperon-hyperon interactions around the hyperon onset density of  $\sim 2n_0$ and beyond. This freedom has been exploited by different groups to build up RMF models that ensure the existence of neutron stars with masses larger than $2 M_\odot$ even with the presence of hyperons (see for example \citep{Weissenborn:2011ut,Bednarek:2011gd,vanDalen:2014mqa,Oertel:2014qza}). While the radii of neutron stars are essentially determined by the nucleonic part of the EoS, the uncertainties in the hyperon interactions reflect on maximum masses that are scattered within a 0.3 $M_\odot$ band, as can be seen from the thorough analysis of various models done by \citep{Fortin:2014mya}, which is consistent with admitting a deviation by at most 30\% of the symmetries assumed to determine the hyperon coupling constants  \citep{Weissenborn:2011ut}. These  results provide an estimate of the uncertainties that one must admit in the hyperonic sector until data on the hyperon interactions at higher densities, coming for instance from HIC experiments \citep{Morita:2014kza}, become available. 
As explained in the preceding paragraph, in this work we have simply made a minimal adjustment of the hyperon parameters away from the symmetry constraints imposed by Eq.~({\ref{eq:couplings}}) in order to reproduce the known hypernuclear properties.
 
\begin{figure}[t]
\begin{center}
\includegraphics[width=0.45\textwidth]{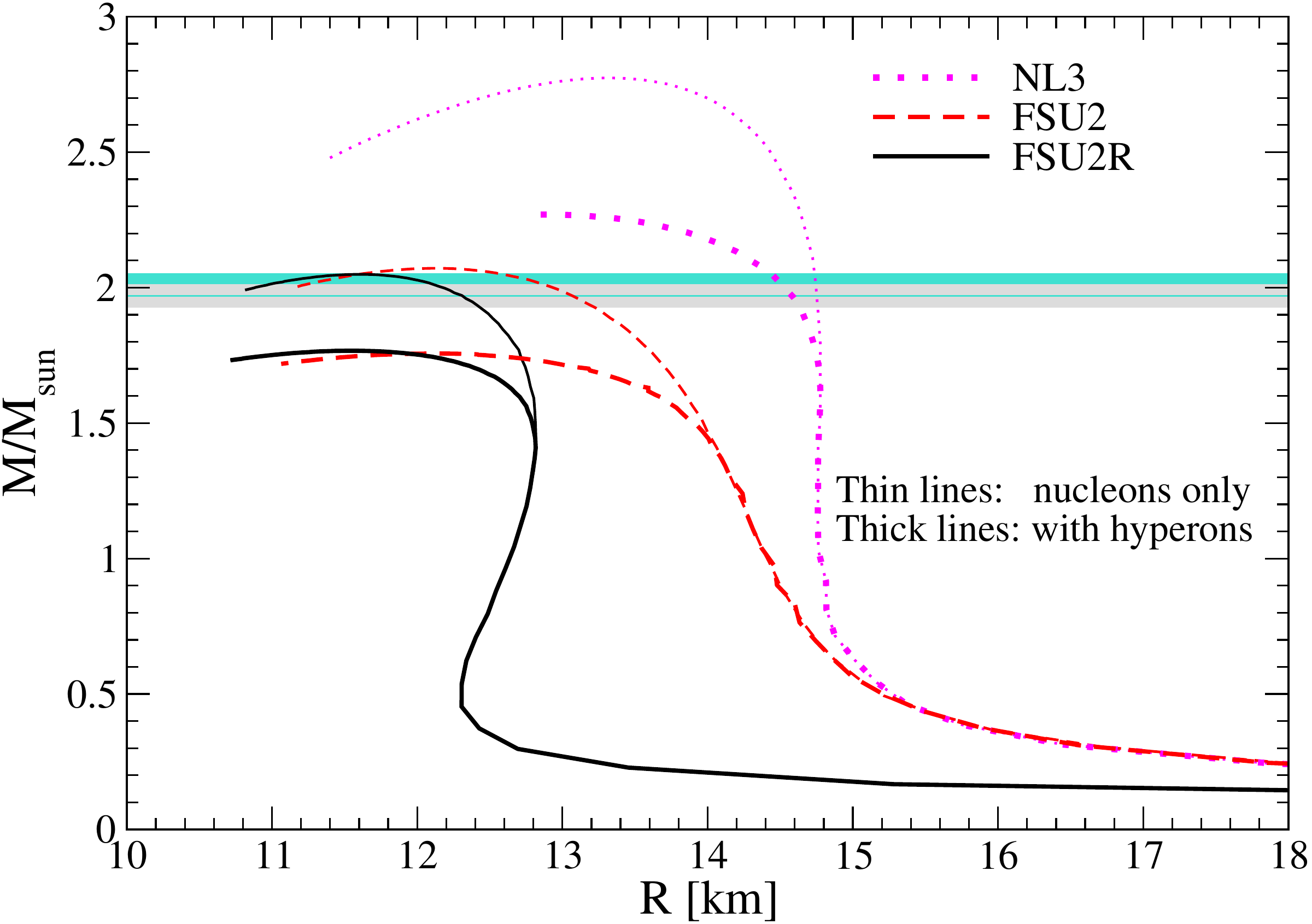}
\caption{Mass versus radius for neutron stars for the models NL3 \citep{Lalazissis:1996rd}, FSU2 
\citep{Chen:2014sca} and FSU2R (this work) with hyperons (thick lines) and 
without hyperons (thin lines). The two shaded bands portray the masses 
$M=1.97 \pm 0.04 M_\odot$ in the pulsar PSR J1614--2230  (gray band) 
\citep{Demorest:2010bx}  and $M=2.01 \pm 0.04 M_\odot$ in the pulsar PSR J0348+0432 
(turquoise band) \citep{Antoniadis:2013pzd}. }
\label{fig:mass-radius-hyperons}
\end{center}
\end{figure}

In Fig.~\ref{fig:mass-radius-hyperons} we show how the presence of hyperons affects the M-R relationship for some representative nuclear EoSs selected from the previous section: the highly stiff EoS of the NL3 model and the FSU2 and FSU2R EoSs. These models differ essentially on the lower density  ($n
\lesssim 2 n_0$) and/or the higher density ($n 
\gtrsim 2 n_0$) stiffness of the EoS.  As already noted in Refs.~\citep{Horowitz:2001ya,Chen:2014sca,Chen:2015zpa} and also discussed in the previous section, models with a larger value of the $\Lambda_\omega$ coupling produce a softer symmetry energy and, in consequence, become more compressible leading to stars with higher central densities and smaller radii. The presence of hyperons softens the EoS by essentially releasing Fermi pressure. Thereby, the stars get further compressed than their nucleonic counterparts, and the maximum masses get reduced by about 15\%,  as seen by the thick lines in Fig.~\ref{fig:mass-radius-hyperons}. 
It is also seen from this figure that the occurrence of hyperons leaves the stellar radii almost unaffected. 

Except for the NL3 model, which has shown to be exceedingly stiff at supranormal densities, the maximum masses of hyperonic stars attained by the other two models are too low, of about $M_{\rm max} = 1.8M_\odot$, to reproduce the 2$M_\odot$ limit. The specific values of the maximum masses of hyperonic stars for all these models can be seen in Table~\ref{tab:starprops} in the '$pneY\mu$'  section. We observe a weak sensitivity to the slope of the symmetry energy, as FSU2 and FSU2R produce essentially the same maximum masses. This was already noted in the context of nucleonic-only EoSs in Ref.~\citep{Horowitz:2001ya}. As the symmetry energy softens, the star simply becomes more compressed and attains a larger central density, but reaches a similar maximum mass value. This phenomenology remains when hyperons are present, as was also found in \citep{Providencia:2012rx}, the only difference being that the hyperonic stars attain a lower maximum mass and have a higher central density than their nucleonic-only counterparts, as expected for a softer EoS. This can  be seen upon comparing the values shown in the  '$pne\mu$' and  '$pneY\mu$' sections of Table~\ref{tab:starprops}.

Since the hyperonic EoS based on the FSU2R model does not produce $M_{\rm max}> 2M_\odot$,  we tense the parameters of this nuclear model a little further so as to make it stiffer. We essentially reduce the value of $\zeta$ from 0.024 to 0.008, which stiffens the EoS at densities larger than twice the saturation density, i.e. around the region where hyperons start appearing (see the hyperon onset density for the different models in Table~\ref{tab:starprops}). The remaining parameters of the model are refitted so as 
to reproduce the  SNM saturation properties of the FSU2
model and a symmetry energy $E_{\rm sym} = 26.2$ MeV at density $n=0.10$~fm$^{-3}$.
The values of the parameters of this new interaction, named  FSU2H, are listed in Table~\ref{t-parameters}, together with the predicted $E_{\rm sym}$ value at saturation density and its slope $L$, which fall comfortably within the newest empirical and theoretical constraints of these quantities, as can be seen in Fig~\ref{fig:esym-l}.  The couplings of the hyperons to the different vector mesons can be readily obtained from Eq.~(\ref{eq:couplings}), and those to the $\sigma$ meson, determined from fixing the hyperon potentials in  SNM, are $g_{\sigma \Lambda}=0.6113$, $g_{\sigma \Sigma}=0.4665$  and $g_{\sigma \Xi}=0.3157$.

We note that the FSU2H interaction produces a certain overpressure in SNM at $n \gtrsim 2 n_0$, since the pressure falls above the allowed region obtained from the modeling of collective flow in HICs, as seen by the long dashed line in the upper part of Fig.~\ref{fig:pressure}. Nevertheless, the EoS for PNM, seen in the lower panel of this figure, falls within the PNM extrapolated band compatible with collective flow.  Since neutron-star matter in beta equilibrium is highly asymmetric we consider this model to be sufficiently realistic to describe neutron stars, whose properties are presented in 
Table~\ref{tab:starprops}. We observe that the maximum mass of $2.38 M_\odot$ obtained for a '$pne\mu$' neutron star with the FSU2H model gets reduced to $2.03 M_\odot$, with a radius of 12 km, when hyperons are present.  We also observe that the radius of a canonical star of  $\sim1.5M_\odot$ gets slightly enhanced from  12.8 km for FSU2R to 13.2 km for FSU2H, which is the price one pays for having stiffened the EoS.

On comparing the '$pne\mu$' with the '$pneY\mu$'  parts of Table~\ref{tab:starprops} we essentially see, as in Fig.~\ref{fig:mass-radius-hyperons}, a reduction of about 15\% on the maximum mass when hyperons are allowed to appear in the neutron star cores. Since the hyperonic EoSs become more compressible, the '$pnYe\mu$' stars attain higher central densities,  but the radii of the maximum-mass stars stay rather similar to their nucleonic-only counterparts. 

We would like to note that the FSU2H parameterization, which produces $M_{\rm max} > 2M_\odot$ even in the presence of hyperons, fulfills  the pressure constraint in neutron star matter at saturation density $n_0$:
\begin{equation}
1.7 \mbox{ MeV fm}^{-3} < P(n_0) < 2.8 \mbox{ MeV fm}^{-3} \ ,
\label{eq:pres_band}
\end{equation}
estimated in \citep{Fortin:2014mya} from the results shown in \citep{Hebeler:2013nza}, which were obtained from microscopic calculations of PNM based on chiral two-nucleon and three-nucleon interactions, and which are in remarkable agreement with the Quantum Monte Carlo
results of Ref.~\citep{Gandolfi:2011xu}, obtained from the phenomenological Argonne v18 NN potential plus three-nucleon forces.
It is argued in  \citep{Fortin:2014mya}  that nearly all hyperonic EoS models that are able to sustain $M_{\rm max} > 2 M_\odot$ produce large PNM pressures of about 5 MeV fm$^{-3}$ 
at saturation density, leading to an overpressure of the nucleonic
(pre-hyperon) segment  and resulting in large radii of around 14 km or more for neutron stars  in the range $1 < M/M_\odot < 1.6$. 
Our FSU2H model does not encounter this problem, since it gives a PNM  pressure of $\sim 2$ MeV fm$^{-3}$ at $n_0$ (see Table \ref{t-parameters}), well within the constraint of Eq.~(\ref{eq:pres_band}), and as a consequence is able to reach a smaller radius of 13 km. We note that the symmetry energy slope parameter $L$ of the hyperonic models analyzed in  \citep{Fortin:2014mya} lies in the range of values $67-118$ MeV, which deviate considerably from the recent constraints displayed in Fig~\ref{fig:esym-l}.

\begin{figure}[ht]
\begin{center}
\includegraphics[width=0.45\textwidth]{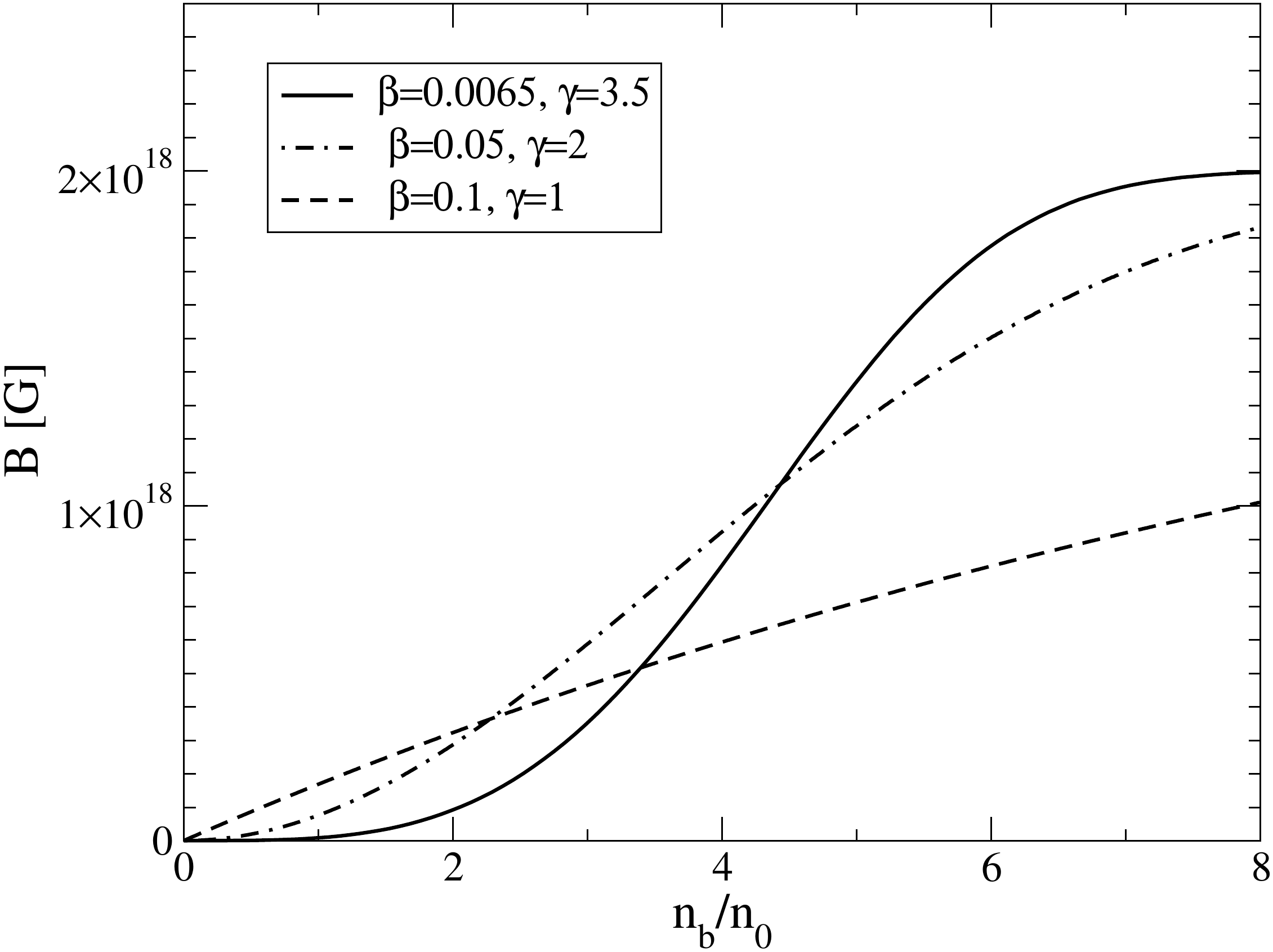}
\caption{Magnetic field versus baryonic density for a function of the type of Eq.~(\ref{eq:B}), taking $B_s=10^{15}$ G and $B_c=2\times 10^{18}$ G, for $(\beta,\gamma)=(0.0065,3.5)$ (solid line),  $(\beta,\gamma)=(0.05,2)$ (dashed-dotted line)   and $(\beta,\gamma)=(0.1,1)$ (dashed line).}
\label{fig:B}
\end{center}
\end{figure}

We now discuss the effect of including a magnetic field in our nucleonic and hyperonic stars. We consider a density-dependent magnetic field with a profile of the type
\begin{equation}
B(n) = B_s + B_c\left\{1 - {\rm exp}\left[-\beta\left(n/n_0\right)^\gamma\right] \right\} \ ,
\label{eq:B}
\end{equation}
introduced in \citep{Chakrabarty:1997ef} and employed in several other works \citep{Rabhi:2009ii,Sinha:2010fm,Lopes:2012nf}. We take a surface magnetic field value of $B_s=10^{15}$ G, consistent with the surface magnetic fields of observed magnetars \citep{Vasisht:1997je,Kouveliotou:1998ze,Woods:1999wa} and a core magnetic field value of $B_c=2\times 10^{18}$ G, which is sufficiently strong to produce distinguishable effects on the properties of neutron stars. The parameters $\beta$ and $\gamma$ control the density where the magnetic field saturates and the steepness of the transition from the surface to the core field, respectively. We take $\beta=0.0065$ and $\gamma=3.5$ which ensure that the magnetic field has practically saturated to its maximum value at around $5-6n_0$, a range that covers the typical central densities of the maximum mass neutron stars explored in this work. Moreover, the indicated $\beta$ and $\gamma$ values produce moderate field values below saturation density, as can be seen by the solid line in Fig.~\ref{fig:B}. We note that this field profile does not incur on instabilities of the parallel component of the pressure $P_{||}$ associated to rapidly rising magnetic field toward relatively strong central values \citep{Sinha:2010fm}.

\begin{figure}[ht]
\begin{center}
\includegraphics[width=0.45\textwidth,]{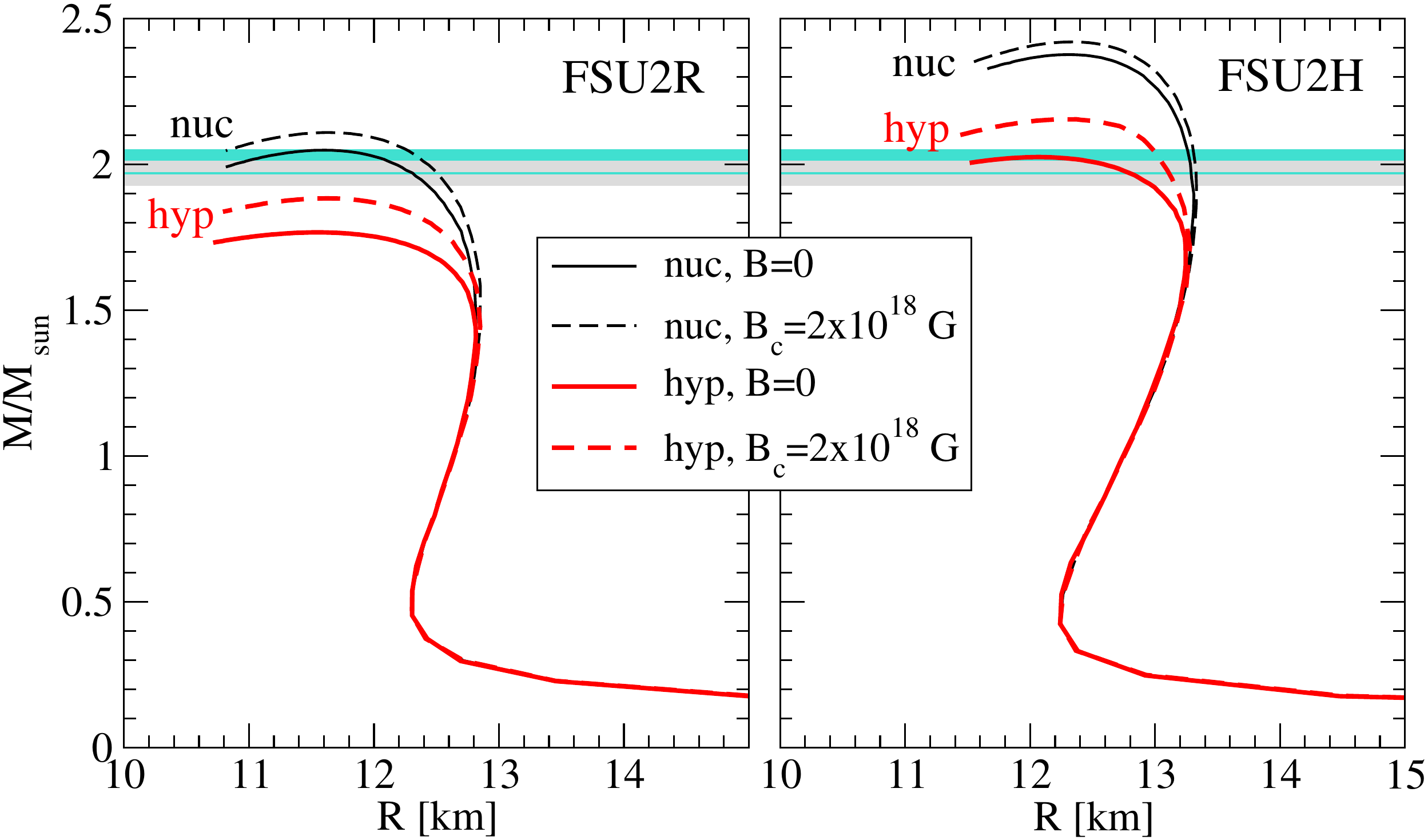}
\caption{Mass versus radius of neutron stars for FSU2R (left panel) and FSU2H (right 
panel) models, with (thick lines) or without (thin lines) hyperons, and without 
(solid lines) or with (dashed lines) a magnetic field with the profile of 
Fig.~\ref{fig:B}, for $\beta=0.0065$ and $\gamma=3.5$ The two shaded bands portray 
the masses $M=1.97 \pm 0.04 M_\odot$ in the pulsar PSR J1614--2230  (gray 
band) \citep{Demorest:2010bx} and $M=2.01 \pm 0.04 M_\odot$ in the pulsar PSR 
J0348+0432 (turquoise band) \citep{Antoniadis:2013pzd}.  }
\label{fig:mass-radius-hyp-b}
\end{center}
\end{figure}

The effect of this magnetic field on the M-R relationship
is displayed in Fig.~\ref{fig:mass-radius-hyp-b}. On the left (right) panel we show the results obtained for the EoS employing the FSU2R (FSU2H) model.  The solid lines correspond to vanishing magnetic field, while the dashed lines include the effects of  the magnetic field with the density profile discussed above.  The thin black lines show the results for nucleonic neutron stars and the thick red lines correspond to the hyperonic stars. 
As observed in earlier works \citep{Lopes:2012nf}, including the magnetic field produces stars with larger maximum masses. This is essentially a consequence of the increase in the total pressure which, apart from the matter pressure $P_{\rm matt}$, also includes the extra average field pressure component, as seen in Eq.~(\ref{press}).  The size of this enhancement is larger for the hyperonic than for the nucleonic stars, which is essentially due, as we will show below, to the additional effect of de-hyperonization that takes place in the presence of a magnetic field. The reduction of hyperons is responsible for enhancing the value of the matter pressure $P_{\rm matt}$, since the Fermi contributions of the other species are larger than in the $B=0$ case.  Nevertheless,  the increase in the maximum mass induced by magnetic field effects is not enough to produce hyperonic star masses of the order of 2$M_{\odot}$ in the case of the FSU2R model, as the dashed red line on the left panel does not reach the observational bands. The effects of the magnetic field on the mass-radius relationships obtained with the FSU2H EoS (right panel) are similar to those for the FSU2R EoS, the only difference being that the constraint $M_{\rm max} > 2 M_\odot$ is now amply fulfilled, since the FSU2H model served this purpose even in the absence of a magnetic field.

We now explore the effect of employing different magnetic field profiles having the same surface and central values, $B_s=10^{15}$ G and $B_c=2\times10^{18}$ G, but different $\beta$ and $\gamma$ parameters. To this end, we consider, in addition to the profile obtained with the parameters $(\beta,\gamma)=(0.0065,3.5)$ chosen in this work, the profiles with $(\beta,\gamma)=(0.05,2)$  \citep{Rabhi:2009ii}  and $(\beta,\gamma)=(0.1,1)$ \citep{Sinha:2010fm}, which are represented, respectively, by the dashed-dotted and dashed lines in Fig.~\ref{fig:B}. We observe that our parameterization produces a substantially lower magnetic field in the $n< 2n_0$ region and reaches 90\% of the saturation value around 5$n_0$, while the dashed-dotted parametrization does it right after $6n_0$. The parameterization of the dashed line does not even reach the value $B=10^{18}$ G within the densities of interest ($n\lesssim 6n_0$).

\begin{figure}[ht]
\begin{center}
\includegraphics[width=0.40\textwidth]{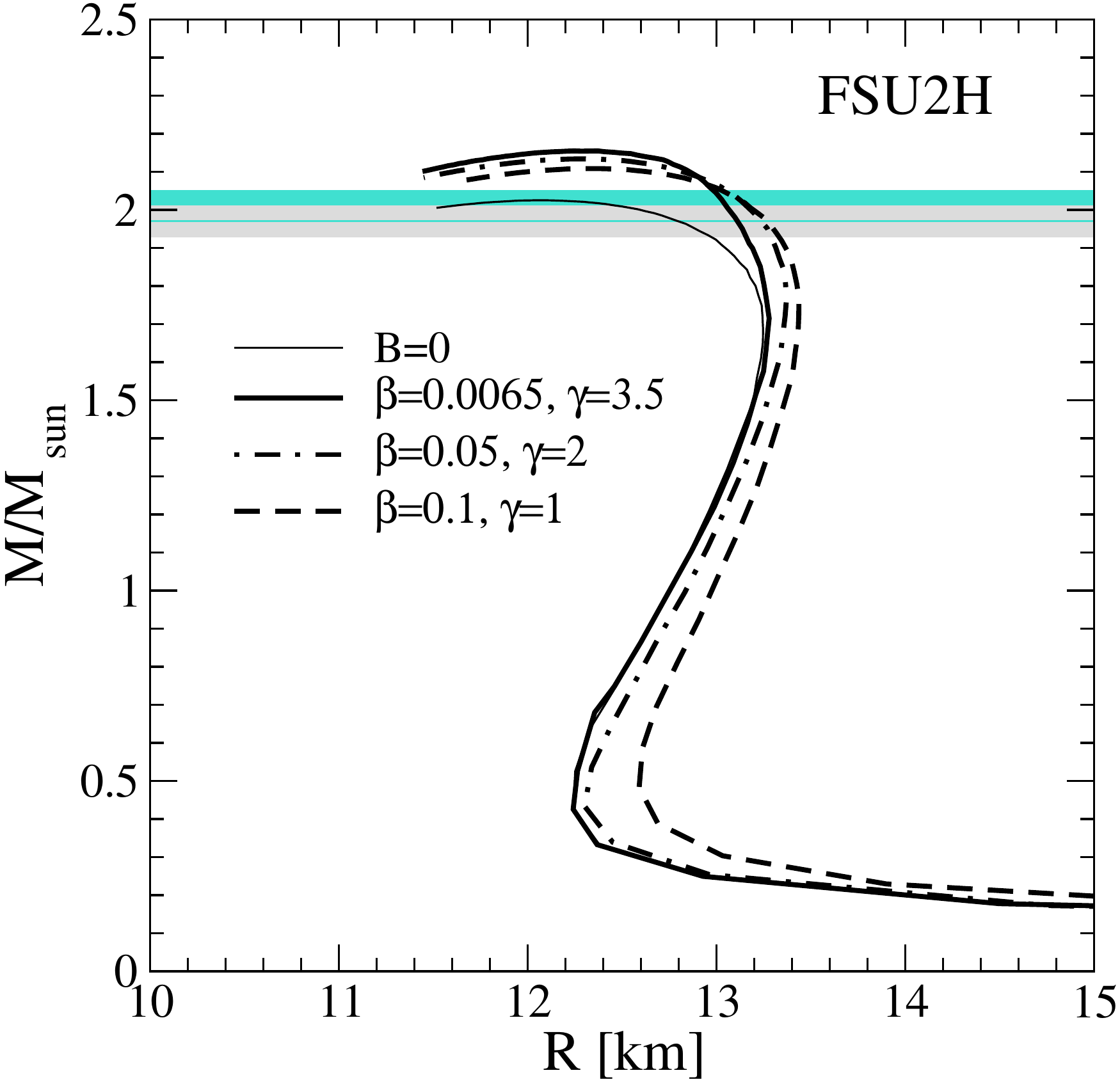}
\caption{Mass versus radius of hyperonic neutron stars obtained with the FSU2H model and 
including a magnetic field with the different profiles displayed in 
Fig.~\ref{fig:B}. The field-free case is shown by the thin solid line. The two 
shaded bands portray the masses $M=1.97 \pm 0.04 M_\odot$ in the 
pulsar PSR J1614--2230 (gray band) \citep{Demorest:2010bx} and $M=2.01 \pm 0.04 
M_\odot$ in the pulsar PSR J0348+0432 (turquoise band) \citep{Antoniadis:2013pzd}. }
\label{fig:mass-radius-hyp-bprofile}
\end{center}
\end{figure}

In Fig.~\ref{fig:mass-radius-hyp-bprofile} we display the M-R relationships obtained with these profiles, together with the zero magnetic field case, represented by a thin solid line. A noticeable dependence of the M-R relationship on the magnetic field profile is observed. The results for the $(\beta,\gamma)=(0.05,2)$ case (thick dashed-dotted line)  are similar to those obtained with our $(\beta,\gamma)=(0.0065,3.5)$ parameterization (thick solid line), but the stars are produced with a somewhat larger radius since the magnetic field and, hence, the total pressure are larger in the pre-hyperon region. This is even more evident  for the M-R relationship obtained with the $(\beta,\gamma)=(0.1,1)$ profile (thick dashed line), which produces stars that are $\sim 0.5$ km wider than the other two cases and deviates from the 13 km maximum radius constraint. The reason is that this profile clearly gives larger magnetic fields in the $n \lesssim n_0$ region, hence producing a larger total pressure and making the star less compressible.

\begin{figure}[ht]
\begin{center}
\includegraphics[width=0.45\textwidth]{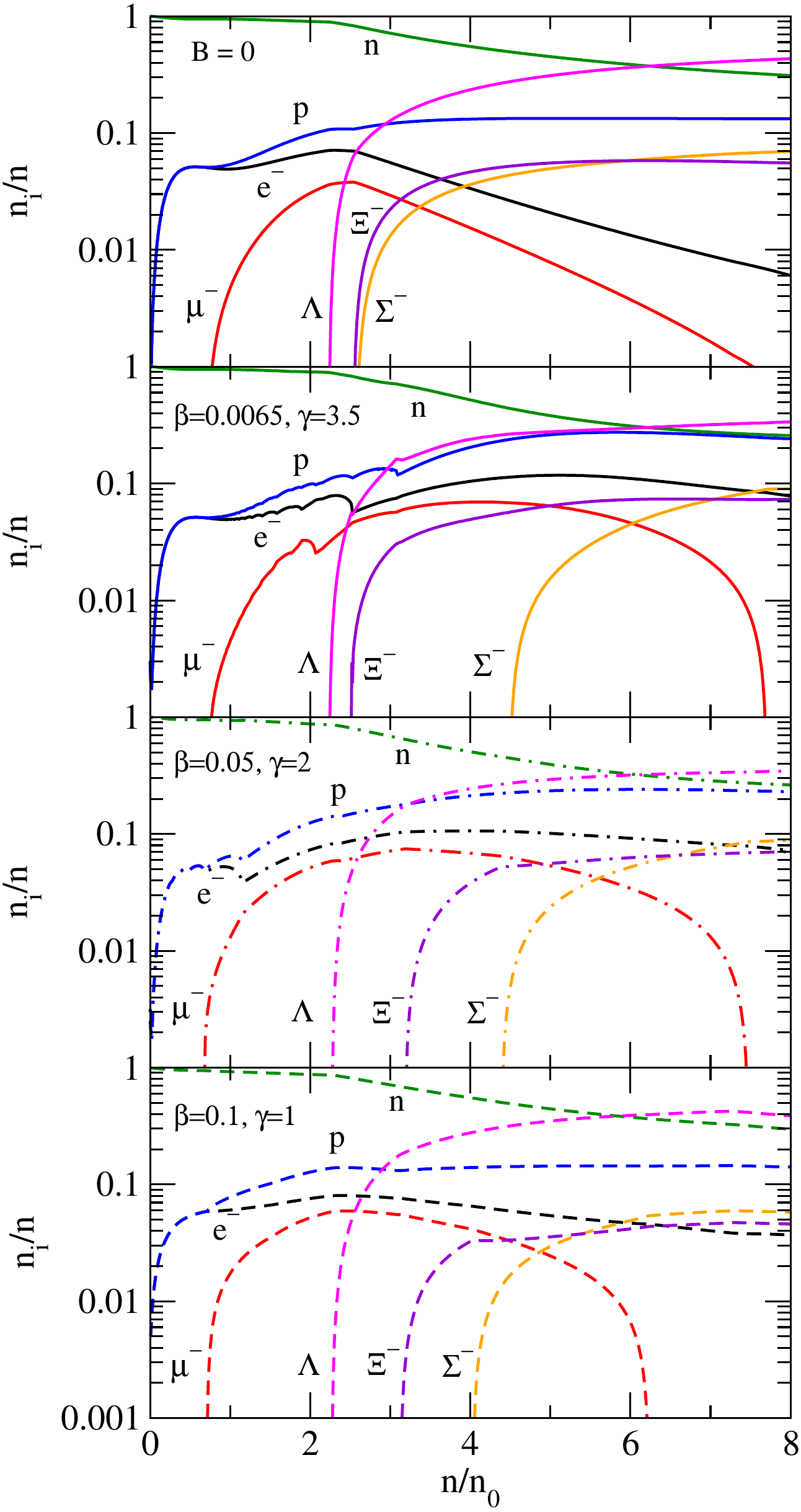}
\caption{Particle fractions as functions of the baryonic density for the FSU2H model without  magnetic field (first panel) and including the magnetic field profile of Eq.~(\ref{eq:B}), taking $B_s=10^{15}$ G and $B_c=2\times10^{18}$ G and for $(\beta,\gamma)=(0.0065,3.5)$ (second panel) , $(\beta,\gamma)=(0.05,2)$ (third panel) and $(\beta,\gamma)=(0.1,1)$ (fourth panel).}
\label{fig:fractions}
\end{center}
\end{figure}

The particle fractions for beta-stable neutron star matter obtained using the FSU2H EoS are shown in Fig.~\ref{fig:fractions} as functions of the baryonic density. The upper panel displays the fractions in the absence of magnetic field, 
while the other panels implement the magnetic field with the three different profiles shown in Fig.~\ref{fig:B}.
Landau oscillations are seen in the charged particle fractions when a magnetic field is applied, reflecting the successive filling of the Landau levels as the quantity $(E_F^2-{m^*}^2)/2|q| B$ reaches integer values. For a fixed density, smaller magnetic fields accommodate more Landau levels and, correspondingly, more oscillations are observed, as seen for instance when comparing the three $B\neq 0$ panels in the $n<2n_0$ density region, where the smallest field corresponds to the $(\beta,\gamma)=(0.0065,3.5)$ case. As density increases, so does the magnetic field in all the considered profiles, eventually needing only one Landau level to accommodate the population of the charged particles. The oscillations then tend to smooth out and disappear with increasing density.
 As is evident, the magnetic field mostly affects the charged particles, which in general increase their population with respect the $B=0$ case. At low and intermediate densities up to $n\sim 4n_0$, we clearly observe an increase in the occupation of negatively charged electrons and muons. This delays the appearance of the negatively charged hyperons, an effect that is especially visible for the $\Sigma^-$ baryon, whose onset moves to $n \gtrsim 4 n_0$ for all the considered magnetic field profiles.

According to the results shown in the second panel of Fig.~\ref{fig:fractions}, in the case of a magnetized hyperonic star having a mass of about $2M_\odot$ with a maximum density of about $5n_0$ (see  Table~\ref{tab:starprops}), the baryon fractions at the center would be 38\% for n, 28\% for $\Lambda$, 26\% for p, 6\% for $\Xi^-$ and 2\% for $\Sigma^-$. In the $B=0$ case (upper panel), these fractions would be 45\% for n, 31\% for $\Lambda$, 13\% for p, 6\% for $\Xi^-$ and 5\% for $\Sigma^-$. We then see that the proton abundance can be twice as large in a magnetar as it is in a field-free star. Our results are qualitatively consistent with those obtained by other works in the literature studying the effect of magnetic fields in hyperonic stars \citep{Broderick:2001qw,Rabhi:2009ii,Sinha:2010fm,Lopes:2012nf,Yue:2009jh}.
We can conclude that, in general, hyperonic magnetars re-leptonize and de-hyperonize with respect to zero-field stars, while the proton abundance increases substantially. This might facilitate direct Urca processes, drastically altering the cooling evolution of the star.

\vspace*{5mm}
\section{Summary}
\label{sec:summary}

We have obtained a new EoS for the nucleonic inner core of neutron stars that fullfills  the constraints coming from recent astrophysical observations of maximum masses and determinations of radii, as well as the requirements from experimental nuclear data known from terrestrial laboratories. This EoS results from a new parameterization of the FSU2 force \citep{Chen:2014sca}, the so-called FSU2R model, that reproduces: i) the 2$M_{\odot}$ observations  \citep{Demorest:2010bx,Antoniadis:2013pzd}, ii) the recent determinations of radii below 13 km region  \citep{Guillot:2013wu,Lattimer:2013hma,Heinke:2014xaa,Guillot:2014lla,Ozel:2015fia,Lattimer:2015nhk}, iii) the saturation properties of nuclear matter and finite nuclei \citep{Tsang:2012se,Chen:2014sca} and iv) the constraints extracted from nuclear collective flow \citep{Danielewicz:2002pu} and kaon production \citep{Fuchs:2000kp,Lynch:2009vc} in HICs. 

The FSU2R model is obtained by modifying the $\Lambda_{\omega}$ and $\zeta$ couplings of the Lagrangian simultaneously, while recalculating the couplings $g_{\sigma N}, g_{\omega N}, g_{\rho N}, 
\kappa$, and $\lambda$ to grant the same saturation properties of FSU2 in  SNM and a symmetry energy of 25.7 MeV at $n=0.10$ fm$^{-3}$. On the one hand, radii of 12-13 km are obtained, owing to the fact that we softened the symmetry energy and, consequently, the pressure of PNM at densities $\sim 1.5$-$2 n_0$, while reproducing the properties of nuclear matter and nuclei. Indeed, we obtain  $E_{\rm sym}= 30$ MeV and $L=44$ MeV, which lie within the limits of the recent determinations of Refs.~\citep{Roca-Maza:2015eza,Lattimer:2012xj,Hagen:2015yea}. Moreover, the FSU2R model predicts a neutron skin thickness of 0.133~fm for the $^{208}$Pb nucleus, which is compatible with the recent experimental results  \citep{Abrahamya12,Horowitz:2012tj,Tarbert:2013jze,Roca-Maza:2015eza}. On the other hand, we have stiffened the EoS above twice the saturation density, which satisfies the constraints of HICs \citep{Danielewicz:2002pu,Fuchs:2000kp,Lynch:2009vc} and allows for maximum masses of 2$M_{\odot}$ \citep{Demorest:2010bx,Antoniadis:2013pzd}. All in all, the FSU2R parameterization allows for a compromise between small stellar sizes and large masses, a task that seemed difficult to achieve in up-to-date  RMF models. 

We also analyze the consequences of the appearance of hyperons inside the core of neutron stars.  The values of the hyperon couplings are determined from the available experimental information on hypernuclei, in particular by fitting to the optical potential of hyperons extracted from the data.  On the one hand, we find that the radii of the neutron stars are rather insensitive to the appearance of the hyperons and, thus, still respect the observations of radii $<$ 13 km. On the other hand,  we obtain a reduction of the maximum mass below 2$M_{\odot}$ once hyperons appear due to the expected softening of the EoS.  However, by refitting the parameters of the FSU2R model slightly, the new parameterization FSU2H fulfills the 2$M_{\odot}$ limit while still reproducing the properties of nuclear matter and nuclei. The price to pay is a stiffer EoS in SNM  as compared to the constraint derived from the modeling of HICs. Nonetheless, the HICs estimate in PNM is still satisfied by the FSU2H parametrization  \citep{Danielewicz:2002pu}. 

We finally study the effect of high magnetic fields on the nucleonic and hyperonic EoSs. This is of particular interest for understanding the behavior of highly magnetized neutron stars, the so-called magnetars. Employing magnetic fields with crustal and interior values of  $ \sim 10^{15}$ G and $\sim 10^{18}$ G, respectively, we find EoSs that are stiffer and produce larger maximum-mass stars, while keeping radii in the 12-13 km range, both for nucleonic and hyperonic magnetars, as long as the magnetic field does not reach values larger than about $10^{17}$ G at saturation density.  The particle fractions in the interior of the stars depend on the specific profile of the magnetic field, but the general trend with respect to zero-field stars is that hyperonic magnetars re-leptonize and de-hyperonize, while the amount of protons may double, a fact that may trigger direct Urca processes affecting the cooling and other transport properties of the star.  

\section*{Acknowledgements}
We are most grateful to J. Piekarewicz for 
a careful reading of the manuscript and for valuable comments.
L.T. acknowledges support from the Ram\'on y Cajal research programme,
FPA2013-43425-P Grant from Ministerio de Economia y Competitividad (MINECO) and NewCompstar COST Action MP1304.
M.C. and A.R. acknowledge support from Grant No. FIS2014-54672-P from MINECO, Grant No. 2014SGR-401 from Generalitat de Catalunya, and the project MDM-2014-0369 of ICCUB (Unidad de Excelencia Mar\'{\i}a de Maeztu) from MINECO.
L.T. and A.R. acknowledge support from the Spanish Excellence Network on Hadronic Physics FIS2014-57026-REDT from MINECO.

\bibliography{biblio}

\end{document}